\documentclass[vecphys]{svmult}

\usepackage{makeidx}         
\usepackage{graphicx}        
\usepackage{multicol}        
\usepackage[bottom]{footmisc}

\makeindex             

\begin{document}

\title*{Rupture processes in fiber bundle models}
\author{Per C. Hemmer\and 
Alex Hansen\and
Srutarshi Pradhan}
\institute{ Department of Physics, Norwegian
University of Science and Technology, N-7491 Trondheim, Norway 
\texttt{per.hemmer@ntnu.no}
\texttt{alex.hansen@ntnu.no} 
\texttt{pradhan.srutarshi@ntnu.no}}


%
\maketitle{}
\section{Introduction}
Fiber bundles with statistically
distributed thresholds for breakdown of individual fibers are interesting
models of the statics and dynamics of failures in materials under stress.
They can be analyzed to an extent that is not possible for more complex
materials. During the rupture process in a fiber bundle avalanches,
in which several fibers fail simultaneously, occur. We study by analytic
and numerical methods the statistics of such avalanches, and the breakdown
process for several models of fiber bundles. The models differ primarily
in the way the extra stress caused by a fiber failure is redistributed
among the surviving fibers.

When a rupture\index{rupture} occurs somewhere in an elastic medium, the stress elsewhere
is increased. This may in turn trigger further ruptures, which can
cascade to a final complete breakdown of the material. To describe
or model such breakdown processes\index{breakdown process} in detail for a real material is
difficult, due to the complex interplay of failures and stress redistributions.
Few analytic results are available, so computer simulations is the
main tool (See Refs.\ \cite{HHP-Herrmann}, \cite{HHP-Chakrabarti} and \cite{HHP-Sornette} 
for reviews). Fiber bundle models\index{fiber bundle model}, on the other hand, are characterized
by simple geometry and clear-cut rules for how the stress caused by
a failed element is redistributed on the intact fibers. The attraction
and interest of these models lies in the possibility of obtaining
exact results, thereby providing inspiration and reference systems
for studies of more complicated materials.

In this review we survey theoretical and numerical results for several
models of bundles of $N$ elastic and parallel fibers, clamped at
both ends, with statistically distributed thresholds for breakdown
of individual fibers (Fig.\ 1). The individual thresholds $x_{i}$
are assumed to be independent random variables with the same cumulative
distribution function $P(x)$ and a corresponding density function
$p(x)$: \begin{equation}
\mbox{Prob}(x_{i}<x)=P(x)=\int_{0}^{x}p(u)\; du.\label{HHP-1}\end{equation}

\begin{center}\includegraphics[%
  width=3in,
  height=2in,
  angle=-90]{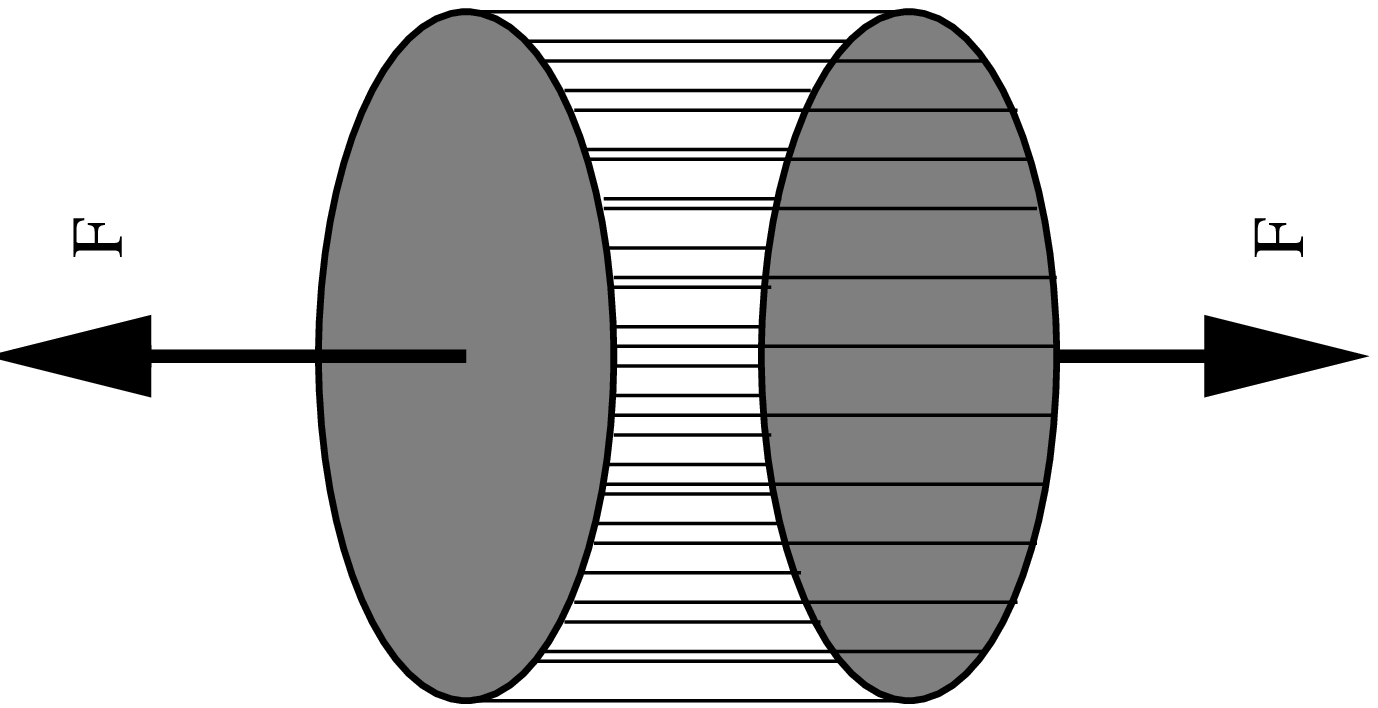}\end{center}

\noindent {\small Fig.\ 1. A fiber bundle of $N$ parallel fibers
clamped at both ends. The externally applied force is $F$.}\\

Whenever a fiber experiences a force equal to or greater than its
strength threshold\index{strength threshold}  $x_{i}$, it breaks immediately and does not contribute
to the strength of the bundle thereafter. The maximal load the bundle
can resist before complete breakdown of the whole bundle is called
the \textit{critical} load. The models differ in the probability distribution\index{probability distribution}
of the thresholds. Two popular examples of threshold distributions
are the uniform distribution\index{uniform distribution} \begin{equation}
P(x)=\left\{ \begin{array}{cl}
x/x_{r} & \mbox{ for }0\leq x\leq x_{r}\\
1 & \mbox{ for }x>x_{r},\end{array}\right.\label{HHP-uniform}\end{equation}
 and the Weibull distribution\index{Weibull distribution} \begin{equation}
P(x)=1-\exp(-(x/x_{r})^{k}).\label{HHP-Weibull}\end{equation}
 Here $x\geq0$, $x_{r}$ is a reference threshold and the dimensionless number
$k$ is the Weibull index (Fig.\ 2).

\includegraphics[%
  width=2.3in,
 height=2.5in]{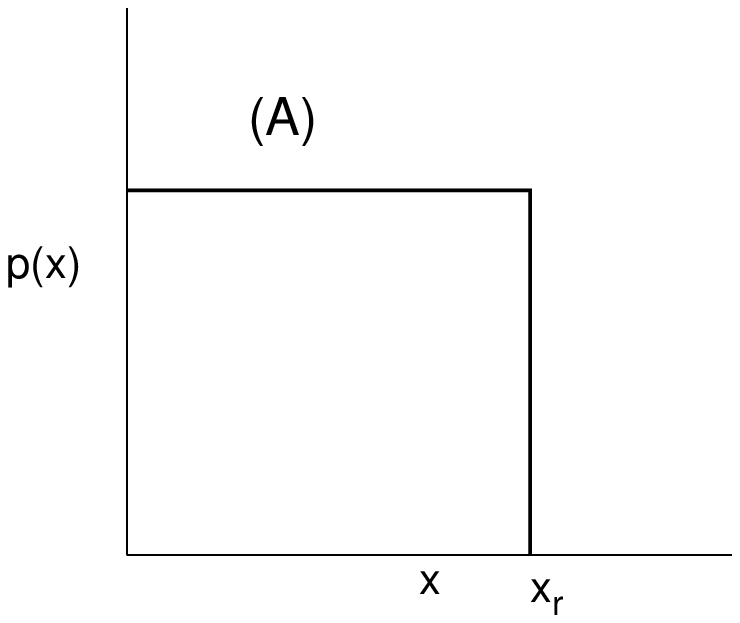}\hskip.1in\includegraphics[%
  width=2.3in,
  height=2.5in]{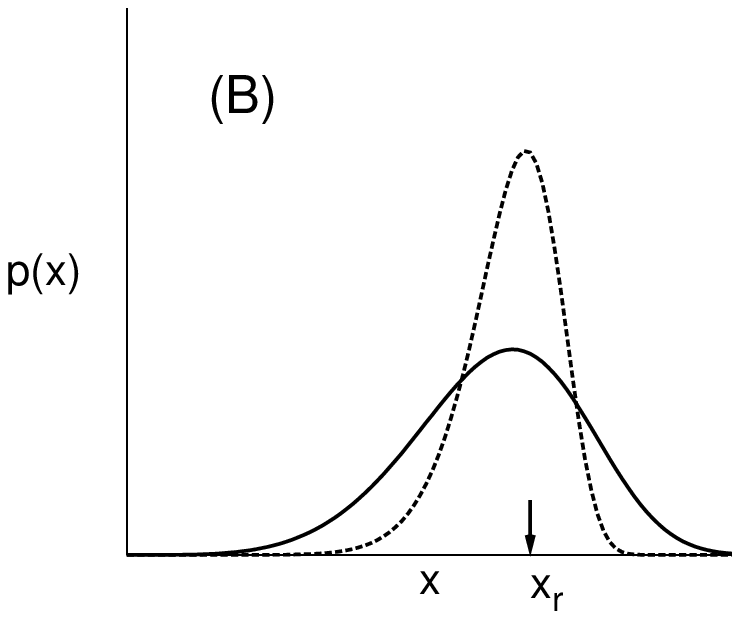}

\vskip.2in

\noindent {\small Fig.\ 2. The uniform distribution (A) and  Weibull
distributions (B) with $k=5$ (solid line) and  $k=10$ (dotted line).}\\

Much more fundamental, however, is the way the models differ in the
mechanism for how the extra stress\index{stress} caused by a fiber failure is redistributed
among the unbroken fibers. The simplest models are the equal-load-sharing\index{equal load sharing}
models, in which the load previously carried by a failed fiber is
shared equally by all the remaining intact bonds in the system. That
some exact results could be extracted for this model was demonstrated
by Daniels \cite{HHP-Daniels} in a classic work some sixty years ago.
Local-load-sharing models\index{local load sharing}, on the other hand, are relevant for materials
in which the load originally carried by a failed fiber is shared by
the surviving fibers in the immediate vicinity of the ruptured fiber.\\

The main property of the fiber bundle breakdown process to be studied
in the present review is the distribution of the sizes of the burst
avalanches\index{avalanche}. The \textit{burst distribution} $D(\Delta)$
is defined as the expected number of bursts in which $\Delta$ fibers
break simultaneously when the bundle is stretched until complete breakdown.
For the equal-load-distribution models that we consider in Sec.\ 2
 Hemmer and Hansen \cite{HHP-Hemmer} showed that  the generic result is a power law, 
\begin{equation}
\lim_{N\rightarrow\infty}\frac{D(\Delta)}{N}\propto\Delta^{-\xi},\label{HHP-powerlaw}\end{equation}
 for large $\Delta$, with $\xi=5/2$. In Sec.\ 2.2 we will show,
however, that for some rather unusual threshold distributions the
power law (\ref{HHP-powerlaw}) is not obeyed. More importantly, we show
in Sec.\ 2.4 that when the whole bundle at the outset is close to
being critical, the exponent $\xi$ crosses over to the value $\xi=3/2$.
In Sec.\ 2.5 we pay particular attention to the rupture process at
criticality, \emph{i.e.}, just before the whole bundle breaks down. 

The average strength of the bundle for a given load can be viewed
as the result of a sequential process. In the first step all fibers
that cannot withstand the applied load break. Then the stress is redistributed
on the surviving fibers, which compels further fibers to burst. This
starts an iterative process that continues until equilibrium is reached,
or all fibers fail. When equilibrium exists, it characterizes the
\emph{average} strength of the bundle for the given load. This recursive
dynamics can be viewed as a fixed-point\index{fixed-point} problem, with interesting
properties when the critical load is approached. We review such  recursive
 dynamics\index{recursive dynamics} in Sec.\ 2.6. 

For other stress redistribution principles than equal-load-sharing,
the avalanche distributions\index{avalanche distribution} are different from the power law (\ref{HHP-powerlaw}).
In Sec.\ 3 we study examples of such systems. Special cases are local-stress-distribution
models in which the surviving nearest neighbors to a failed fiber
share all the extra stress, and a model in which the fibers are anchored
to an elastic clamp.

\section{Equal-load-sharing fiber bundles}

This is the fiber-bundle model with the longest history. It was used by Pierce, in the context of testing the strength of cotton yarn  \cite{HHP-Peirce}.
The basic assumptions are that the fibers obey Hookean elasticity
right up to the breaking point, and that the load distributes itself
\emph{evenly} among the surviving fibers. The model with this democratic
load redistribution is similar to mean-field\index{mean-field} models in statistical physics. 
\begin{center}\includegraphics[%
  width=3in,
  height=2.5in,
  angle=0]{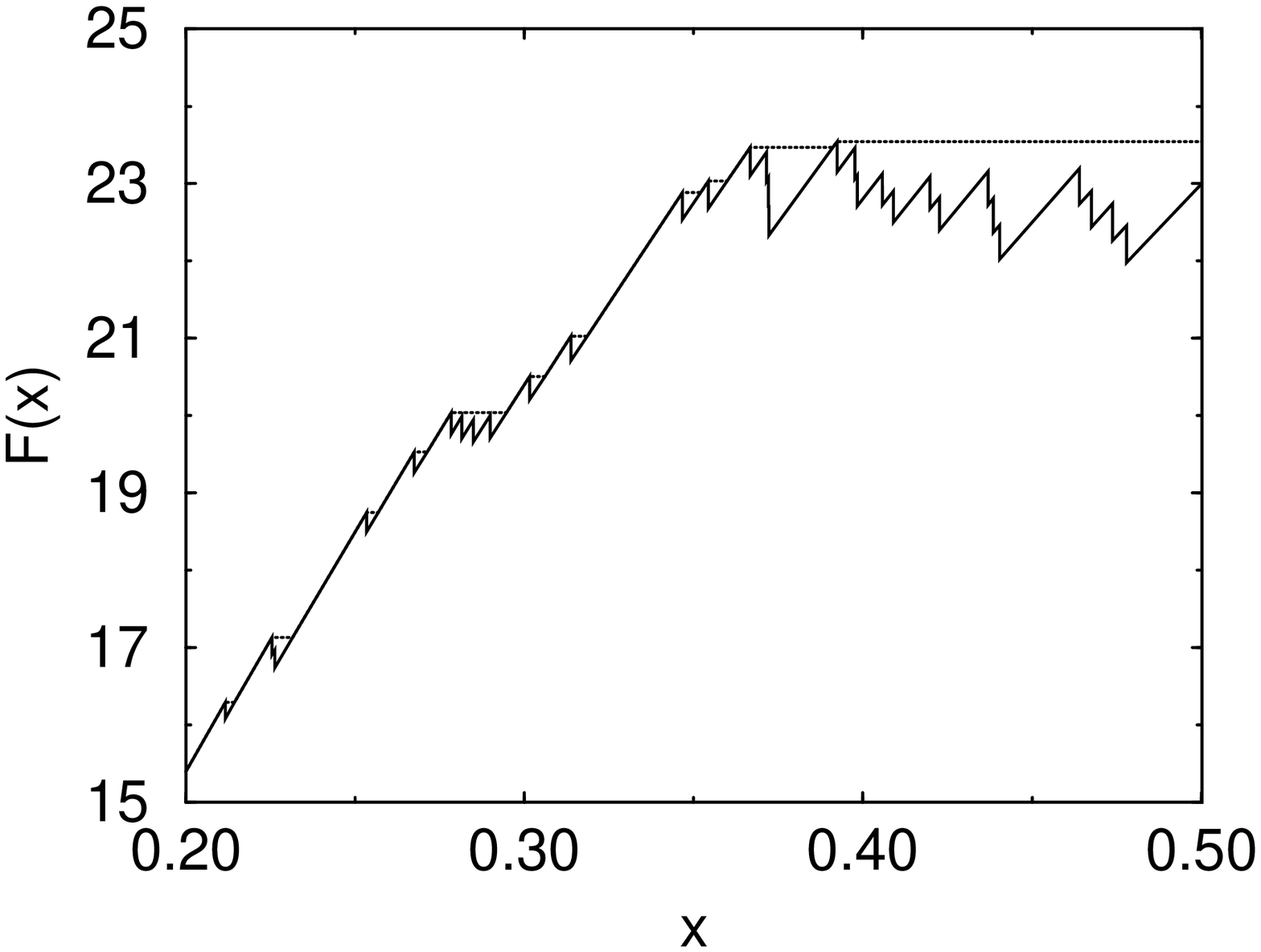}\end{center}

\indent {\small Fig.\ 3. $F(x)$ vs.  $x$ curve. Avalanches are 
shown as horizontal lines.}\\

At a force $x$ per surviving fiber, the total force on the bundle
is \begin{equation}
F(x)=Nx[1-\phi(x)]\;,\label{HHP-3.1}\end{equation}
 where $\phi(x)$ is the fraction of failed fibers. In Fig. 3 we
show an example of a $F$ \textit{vs.} $x$. We have in mind an experiment
in which the force $F$, our control parameter (Fig. 1), is steadily
increasing. This implies that not all parts of the $F(x)$ curve are
physically realized. The experimentally relevant function is \begin{equation}
F_{ph}(x)={\textrm{LMF}}\,\, F(x)\;,\label{HHP-3.2}\end{equation}
 the least monotonic function not less than $F(x)$. A horizontal
part of $F_{ph}(x)$ corresponds to an avalanche, the size of which
is characterized by the number of maxima of $F(x)$ within the corresponding
range of $x$ (Fig. 3). 

It is the fluctuations in $F(x)$ that create avalanches. For a large
sample the fluctuations will be small deviations from the average
macroscopic characteristics $\langle F\rangle$. This \textit{average}
total force is given by \begin{equation}
\langle F\rangle(x)=Nx[1-P(x)].\label{HHP-4}\end{equation}
 Let us for the moment assume that $\langle F\rangle(x)$ has a single
maximum. The maximum corresponds to the value $x=x_{c}$ for which
$d\langle F\rangle/dx$ vanishes. This gives \begin{equation}
1-P(x_{c})-x_{c}\; p(x_{c})=0.\label{HHP-5}\end{equation}
 The threshold $x_{c}$ corresponding to the maximum of $F$ is denoted
the \textit{critical} threshold. Because of fluctuations, however,
the maximum value of the force may actually occur at a slightly different
value of $x$.

\subsection{Burst distribution: The generic case.}

\begin{center}\includegraphics[%
  width=2.0in,
  height=2.2in]{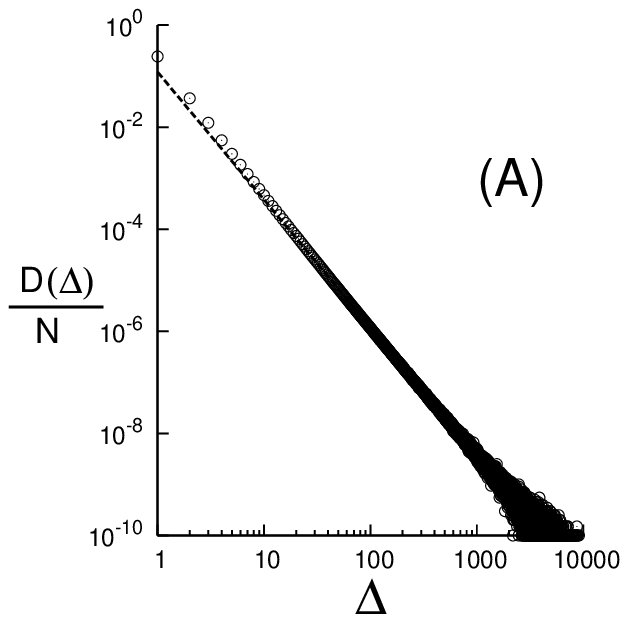}\hskip.2in\includegraphics[%
  width=2.0in,
  height=2.2in]{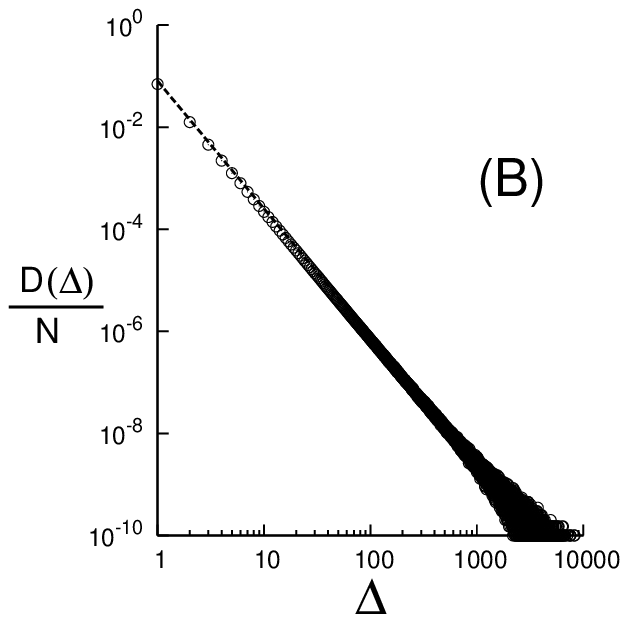}\end{center}

\noindent {\small Fig.\ 4. The burst distribution $D(\Delta)/N$
for the uniform distribution (A) and the Weibull distribution with
index 5 (B). The dotted lines represent the power law with exponent
$-5/2$. Both figures are based on $20000$ samples of bundles each
with $N=10^{6}$ fibers.}\\

That the rupture process produces a power-law decay of the burst distribution
$D(\Delta)$ is seen at once by simulation experiments. Fig. 4 shows
results for the uniform threshold distribution (\ref{HHP-uniform}) and the Weibull
distribution (\ref{HHP-Weibull}) with index $k=5$.
 
In order to derive analytically the burst distribution, let us start
by considering a small threshold interval $(x,x+dx)$ in a range where
the average force $\langle F\rangle(x)$ increases with $x$. For
a large number $N$ of fibers the expected number of surviving fibers
is $N[1-P(x)]$. And the threshold values in the interval, of which there
are $Np(x)dx$, will be Poisson distributed. When $N$ is arbitrarily
large, the burst sizes can be arbitrarily large in any finite interval
of $x$. 

Assume that an infinitesimal increase in the external force results
in a break of a fiber with threshold $x$. Then the load that this
fiber suffered, will be redistributed on the $N[1-P(x)]$ remaining
fibers; thus they experience a load increase \begin{equation}
\delta x=\frac{x}{N[1-P(x)]}.\label{HHP-A1}\end{equation}
 The \emph{average} number of fibers that break as a result of this
load increase is \begin{equation}
a=a(x)=Np(x)\cdot\delta x=\frac{xp(x)}{1-P(x)}.\label{HHP-A2}\end{equation}

For a burst of size $\Delta$ the increase in load per fiber will
be a factor $\Delta$ larger than the quantity (\ref{HHP-A1}), and an
average number $a(x)\Delta$ will break. The probability that precisely
$\Delta-1$ fibers break as a consequence of the first failure is
given by a Poisson distribution\index{Poisson distribution} with this average, i.e.\ it equals
\begin{equation}
\frac{(a\Delta)^{\Delta-1}}{(\Delta-1)!}\; e^{-a\Delta}.\label{HHP-A3}\end{equation}
 This is not sufficient, however. We must ensure that the thresholds
for these $\Delta-1$ fibers are not so high that the avalanche stops
before reaching size $\Delta$. This requires that at least $n$ of
the thresholds are in the interval $(x,x+n\delta x)$, for $1\leq n\leq\Delta-1$.
In other words, if we consider the $\Delta$ intervals $(x,x+\delta x)$,
$(x+\delta x,x+2\delta x)$, \ldots{}, $(x+(\Delta-1)\delta,x+\Delta\delta x)$,
we must find at most $n-1$ thresholds in the $n$ last intervals.
There is the same a priori probability to find a threshold in any
interval. The solution to this combinatorial problem is given in Ref. \cite{HHP-Kloster}.
The resulting probability to find all intermediate thresholds weak
enough equals $1/\Delta$. Combining this with (\ref{HHP-A3}), we have
for the probability $\phi(\Delta,x)$ that the breaking of the first
fiber results in a burst of size $\Delta$: \begin{equation}
\phi(\Delta,x)=\frac{\Delta^{\Delta-1}}{\Delta!}\; a(x)^{\Delta-1}e^{-a(x)\Delta}.\label{HHP-A4}\end{equation}

This gives the probability of a burst of size $\Delta$, as a consequence
of a fiber burst due to an infinitesimal increase in the external
load. However, we still have to ensure that the burst actually \emph{starts}
with the fiber in question and is not part of a larger avalanche starting
with another, weaker, fiber. Let us determine the probability $P_{b}(x)$
that this initial condition is fulfilled. 

For that purpose consider the $d-1$ fibers with the largest thresholds
below $x$. If there is no strength threshold in the interval $(x-\delta x,x)$,
at most one threshold value in the interval $(x-2\delta x,x)$, ...
, at most $d-1$ values in the interval $(x-d\delta x,x)$, then the
fiber bundle can not at any of these previous $x$-values withstand
the external load that forces the fiber with threshold $x$ to break.
The probability that there are precisely $h$ fiber thresholds in
the interval $(x-\delta x\; d,x)$ equals \[
\frac{(ad)^{h}}{h!}\; e^{-ad}.\]
 Dividing the interval into $d$ subintervals each of length $\delta x$,
the probability $p_{h,d}$ that these conditions are fulfilled is
exactly given by $\; p_{h,d}=1-h/d$ (See Ref.\ \cite{HHP-Kloster}).
Summing over the possible values of $h$, we obtain the probability
that the avalanche can not have started with the failure of a fiber
with any of the $d$ nearest-neighbor threshold values below $x$:
\begin{equation}
P_{b}(x|d)=\sum_{h=0}^{d-1}\frac{(ad)^{h}}{h!}\; e^{-ad}(1-\frac{h}{d})=(1-a)e^{-ad}\sum_{h=0}^{d-1}\frac{(ad)^{h}}{h!}\;+\;\frac{(ad)^{d}}{d!}\; e^{-ad}.\label{HHP-Pb}\end{equation}
 Finally we take the limit $d\rightarrow\infty$, for which the last
term vanishes. For $a>1$ the sum must vanish since the left-hand
side of (\ref{HHP-Pb}) is non-negative, while the factor $(1-a)$ is
negative. For $a<1$, on the other hand, we find \begin{equation}
P_{b}(x)=\lim_{d\rightarrow\infty}P_{b}(x|d)=1-a,\end{equation}
 where $a=a(x)$. The physical explanation of the different behavior
for $a>1$ and $a\leq1$ is straightforward: The maximum of the total
force on the bundle occurs at $x_{c}$ for which $a(x_{c})=1$, see
Eqs.\ (\ref{HHP-5}) and (\ref{HHP-A2}), so that $a(x)>1$ corresponds to
$x$ values almost certainly involved in the final catastrophic burst\index{catastrophic failure}.
The region of interest for us is therefore when $a(x)\leq1$, where
avalanches on a microscopic scale occur. This is accordance with what
we found in the beginning of this section, viz.\ that the burst of
a fiber with threshold $x$ leads immediately to a average number
$a(x)$ of additional failures. 

Summing up, we obtain the probability that the fiber with threshold
$x$ is the first fiber in an avalanche\index{avalanche} of size $\Delta$ as the product
\begin{equation}
\Phi(x)=\phi(\Delta,x)P_{b}(x)=\frac{\Delta^{\Delta-1}}{\Delta!}\; a(x)^{\Delta-1}e^{-a(x)\Delta}[1-a(x)],\label{HHP-prob}\end{equation}
 where $a(x)$ is given by Eq.\ (\ref{HHP-A2}), \[
a(x)=\frac{x\; p(x)}{1-P(x)}.\]

Since the number of fibers with threshold values in $(x,x+\delta x)$ is
$N\; p(x)\; dx$, the burst distribution is given by \begin{equation}
\frac{D(\Delta)}{N}=\frac{1}{N}\int_{0}^{x_{c}}\Phi(x)p(x)\; dx=\frac{\Delta^{\Delta-1}}{\Delta!}\int_{0}^{x_{c}}a(x)^{\Delta-1}e^{-a(x)\Delta}\left[1-a(x)\right]\; p(x)\; dx.\label{HHP-Delta}\end{equation}

For large $\Delta$ the maximum contribution to the integral comes
from the neighborhood of the upper integration limit, since $a(x)\; e^{-a(x)}$
is maximal for $a(x)=1$, i.e.\ for $x=x_{c}$. Expansion around
the saddle point, using \begin{equation}
a^{\Delta}e^{-a\Delta}=\exp\left[\Delta(-1-{\textstyle \frac{1}{2}}(1-a)^{2}+{\mathcal{O}}(1-a)^{3})\right],\end{equation}
 as well as $a(x)\simeq1+a'(x_{c})(x-x_{c})$, produces \begin{equation}
\frac{D(\Delta)}{N}=\frac{\Delta^{\Delta-1}e^{-\Delta}}{\Delta!}a'(x_{c})\int_{0}^{x_{c}}p(x_{c})\;e^{-a'(x_{c})^{2}(x_{c}-x)^{2}\Delta/2}(x-x_{c})\; dx.\label{HHP-saddle}\end{equation}

The integration yields the asymptotic behavior \begin{equation}
D(\Delta)/N\propto\Delta^{-\frac{5}{2}},\label{HHP-asymp}\end{equation}
 universal for those threshold distributions for which the assumption
of a single maximum of $\langle F\rangle(x)$ is valid. 

Note that if the experiment had been stopped before complete breakdown,
at a force per fiber $x$ less than $x_{c}$, the asymptotic behavior
would have been dominated by an \emph{exponential} fall-off rather
than a power law: \begin{equation}
D(\Delta)/N\propto\Delta^{-\frac{5}{2}}e^{-[a(x)-1-\ln a(x)]\Delta}.\end{equation}

When $x$ is close to $x_c$ the exponent is proportional to $(x_c-x)^2\Delta $. The burst distribution then takes the scaling form
 \begin{equation}
 D(\Delta) \propto \Delta^{-\eta}\; G\left(\Delta^{\nu}(x_c-x)\right),
 \end{equation}
 with a Gaussian function $G$, a power law index $\eta =
 \frac{5}{2}$ and $\nu = \frac{1}{2}$. Thus the breakdown process
 is similar to critical phenomena with a critical point at total
 breakdown \cite{HHP-Hemmer,HHP-Hansen2,HHP-Hansen} .


\subsection{Burst distribution: Nongeneric cases}

What happens when the average strength curve, $\langle F\rangle(x)$,
does \emph{not} have a unique maximum? There are two possibilities:
(i) it has \textit{several} parabolic maxima, or (ii) it has \textit{no}
parabolic maxima. 

When there are several parabolic maxima, and the absolute maximum
does not come first (i.e.\ at the lowest $x$ value), then there
will be several avalanche series each terminating at a local critical
point with an accompanying burst of macroscopic size, while the breakdown
of the bundle occurs when the absolute maximum is reached \cite{HHP-Lee}.
The power law asymptotics (\ref{HHP-asymp}) of the avalanche distribution
is thereby unaffected, however. 

The second possibility, that the average strength curve has no parabolic
maxima, is more interesting. We present here two model examples of
such threshold distributions. 

The threshold distribution\index{threshold distribution} for model I is, in dimensionless units,
\begin{equation}
P(x)=\left\{ \begin{array}{cl}
0 & \mbox{ for }x\leq2\\
1-(x-1)^{-1/2} & \mbox{ for }x>2,\end{array}\right.\end{equation}
 while model II corresponds to \begin{equation}
P(x)=\left\{ \begin{array}{cl}
0 & \mbox{ for }x\leq1\\
1-x^{-\alpha} & \mbox{ for }x>1,\end{array}\right.\end{equation}
 with a positive parameter $\alpha$.

\begin{center}\includegraphics[%
  width=2.5in,
  height=2.2in]{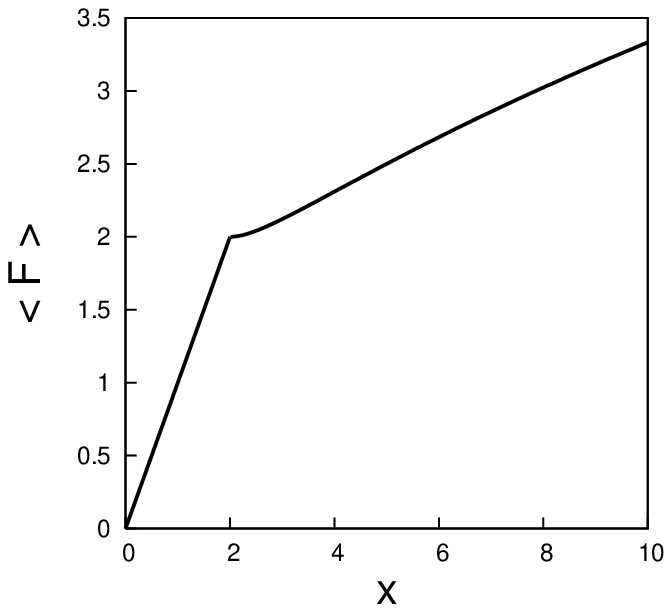}\includegraphics[%
  width=2.5in,
  height=2.2in]{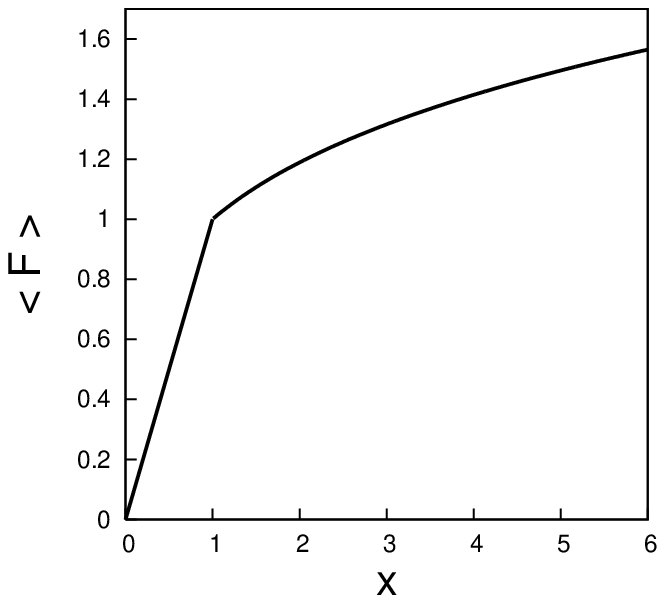}\end{center}

\begin{center}\includegraphics[%
  width=2.5in,
  height=2.2in]{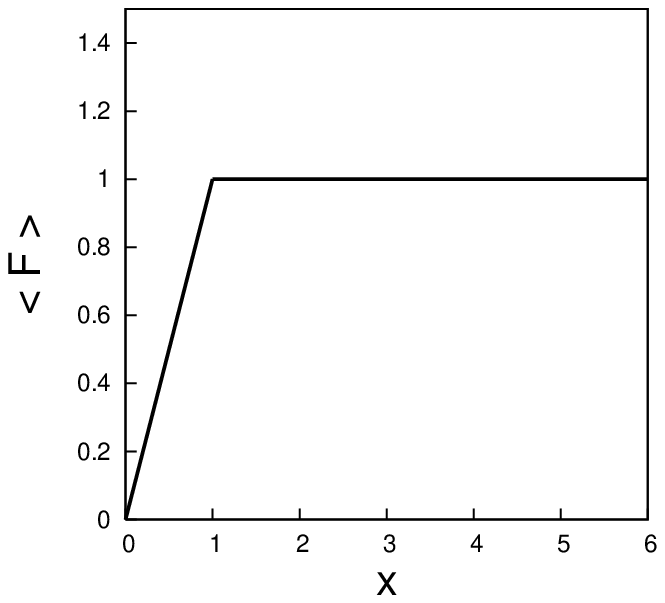}\end{center}

\noindent {\small Fig.\ 5. Average force on the fiber bundle for model
I (upper left), model II for $\alpha=3/4$ (upper right), and model
II in the limit $\alpha\rightarrow1$ (bottom figure).}\\

For the two models the corresponding macroscopic bundle strength per
fiber is, according to Eq.\ (\ref{HHP-Delta}), \begin{equation}
\frac{\langle F\rangle}{N}=\left\{ \begin{array}{cl}
x & \mbox{for }x\leq2\\
\frac{x}{\sqrt{x-1}} & \mbox{for }x>2\end{array}\right.\label{HHP-modelI}\end{equation}
 for model I, and 

\begin{equation}
\frac{\langle F\rangle}{N}=\left\{ \begin{array}{cl}
x & \mbox{for }x\leq1\\
x^{1-\alpha} & \mbox{for }x>1\end{array}\right.\label{HHP-alpha}\end{equation}
 for model II (see Fig.\ 5).\\
 
To calculate the avalanche distribution\index{avalanche distribution} we use (\ref{HHP-Delta}), in
both cases with $x_{c}=\infty$ as the upper limit in the integration.
For Model I the $\langle F\rangle$ graph has at $x=2+$ an extremum,
viz.\ a minimum. At the minimum we have $a(x)=\frac{x}{2(x-1)}=1$.
Since the $\Delta$-dependent factor $a^{\Delta}e^{-a\Delta}$ in
the integrand of (\ref{HHP-Delta}) has a maximum for $a=1$, we obtain
\begin{equation}
D(\Delta)/N\propto\Delta^{-5/2}\end{equation}
 for large $\Delta$. Even if the macroscopic load curve does
not have a maximum at any finite $x$ in this case, the generic power
law\index{power-law} (\ref{HHP-asymp}) holds. 

For model II equation (\ref{HHP-Delta}) gives \begin{equation}
\frac{D(\Delta)}{N}=\frac{1-\alpha}{\alpha}\;\frac{\Delta^{\Delta-1}}{\Delta!}\left[\alpha e^{-\alpha}\right]^{\Delta}\propto\Delta^{-\frac{3}{2}}\left[\alpha e^{1-\alpha}\right]^{\Delta}.\label{HHP-one}\end{equation}
 For $\alpha=3/4$ as in Fig.\ 5, or more generally $\alpha<1$,
the avalanche distribution\index{avalanche distribution} does \emph{not} follow a power law, but
has an exponential cut-off in addition to a $\Delta^{-3/2}$ dependence. 

When $\alpha\rightarrow1$, the average force (\ref{HHP-alpha}) approaches
a constant for $x>1$, and the burst distribution (\ref{HHP-one}) approaches
a power law \begin{equation}
\frac{D(\Delta)}{N}\propto\Delta^{-3/2},\label{HHP-3/2}\end{equation}
 a result easily verified by simulation on the system with $P(x)=1-1/x$  
 for $x\geq1$.
That a power law with exponent $3/2$, different from the generic
burst distribution (\ref{HHP-asymp}), appears when \begin{equation}
\frac{d}{dx}\langle F\rangle\rightarrow0,\end{equation}
 will be apparent when we in Sec.\ 2.5 study what happens at criticality.

\subsection{Mapping onto a random walk problem}

Let $F_{k}$ be the force on the bundle when the $k$th fiber fails.
It is the nonmonotonicities in the sequence $F_{1},F_{2},\ldots$
that produce avalanches of size $\Delta>1$. Let us consider the probability
distribution of the force increase $\Delta F=F_{k+1}-F_{k}$ between
two consecutive bursts, the first taking place at a force $x=x_{k}$
per fiber, so that $F_{k}=(N-k+1)x$. 

Since $\Delta F=(N-k)(x_{k+1}-x)-x$, it follows that \begin{equation}
\Delta F\geq-x.\end{equation}
 The probability to find the $k+1$th threshold in the interval $(x_{k+1},x_{k+1}+dx_{k+1})$,
for given $x=x_{k}$, equals \begin{equation}
(N-k-1)\frac{[1-P(x_{k+1})]^{N-k-2}}{[1-P(x)]^{N-k-1}}\; p(x_{k+1})\; dx_{k+1}.\end{equation}
 By use of the connection $x_{k+1}=x+(\Delta F+x)/(N-k)$ this probability
density for $x_{k+1}$ is turned into the probability density $\rho(\Delta F)\, d\Delta F$
for $\Delta F$: \begin{equation}
\rho(\Delta F)=\frac{N-k-1}{N-k}\;\frac{[1-P(x+(\Delta F+x)/(N-k))]^{N-k-2}}{[1-P(x)]^{N-k-1}}p\left(x+\frac{\Delta F+x}{N-k}\right),\end{equation}
 which is properly normalized to unity. For large $N-k$ this simplifies
to \begin{equation}
\rho(\Delta F)=\left\{ \begin{array}{cl}
0 & \mbox{ for }\Delta F<-x\\
\frac{p(x)}{1-P(x)}\;\exp\left[-\frac{(\Delta F+x)}{1-P(x)}\; p(x)\right] & \mbox{ for }\Delta F\geq-x.\end{array}\right.\label{HHP-step}\end{equation}

\vskip.6in
\begin{center}\includegraphics[%
  width=2in,
  height=1.5in]{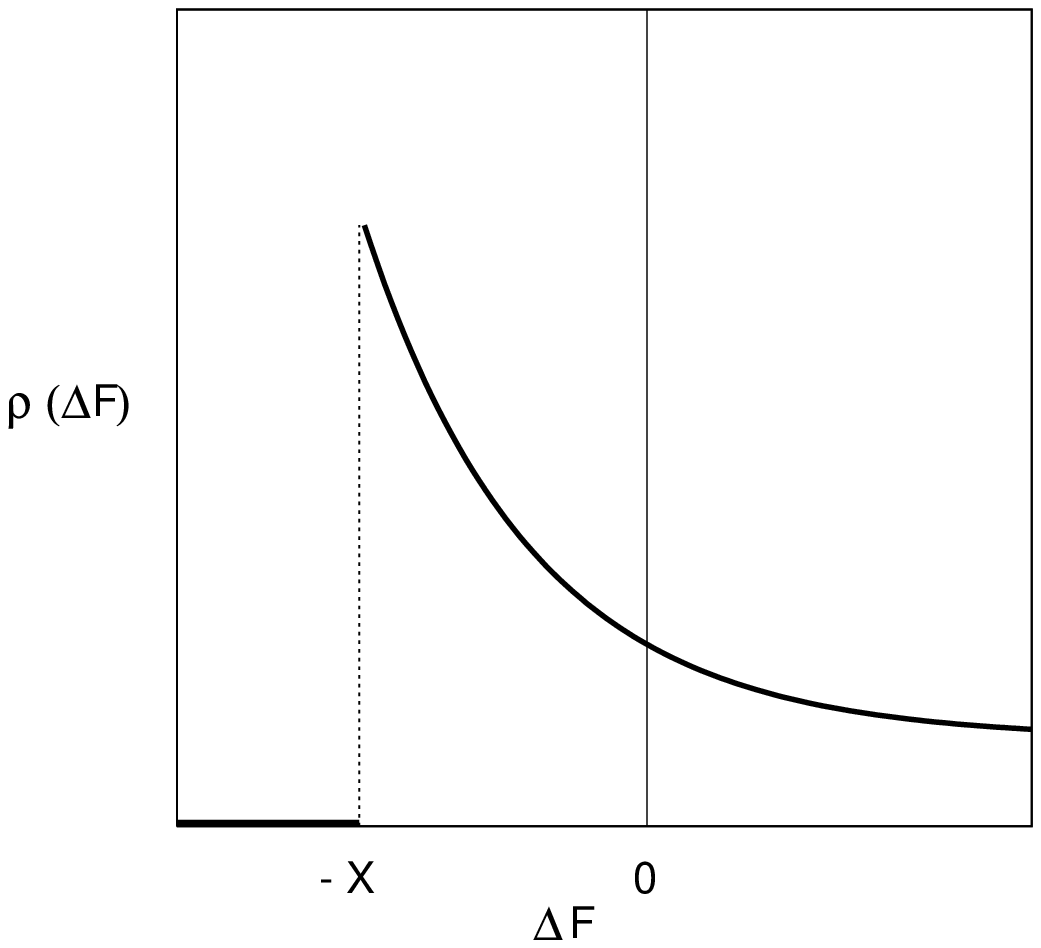}\end{center}
\vskip.2in

\noindent {\small Fig.\ 6. The probability distribution $\rho(\Delta F)$
of the step length in the random walk.}\\
\vskip.1in
The values $F_{1},F_{2},F_{3}\ldots$ of the force on the bundle can
be considered as the positions of a random walker\index{random walk} with the probability
$\rho(\Delta F)$ for the length of the next step \cite{HHP-Hansen2}.
It is a random walk of an unusual type: The step length is variable,
with the steps in the negative direction are limited in size (Fig.\ 6).
In general the walk is \emph{biased} since \begin{equation}
\langle\Delta F\rangle=\int\Delta F\;\rho(\Delta F)\; d\Delta F=\frac{1-P(x)-xp(x)}{p(x)}\label{HHP-stepaverage}\end{equation}
 is zero, \emph{e.g.} unbiased, \emph{only} at the critical threshold
$x_{c}$, given by Eq.\ (\ref{HHP-5}). That the random walk is unbiased
at criticality is to be expected, of course, since the average bundle
strength $\langle F\rangle$ as function of $x$ is stationary here. 

The probability that $\Delta F$ is positive equals \begin{equation}
\mbox{Prob}(\Delta F>0)=\int_{0}^{\infty}\rho(\Delta F)\; d\Delta F=\exp\left[-\frac{xp(x)}{1-P(x)}\right].\label{HHP-positivestep}\end{equation}
 That $\Delta F$ is positive implies that the burst has the length
$\Delta=1$. The result (\ref{HHP-positivestep}) is identical to the
previously determined probability $\phi(1,x)$, (\ref{HHP-A4}), for a
burst of length 1, when we have not ensured that the burst actually
\emph{starts} with the fiber in question and is not part of a larger
avalanche\index{avalanche}. 

In section 2.5 we will see that the random-walk analogy can be used
in a quantitative way to predict the avalanche distribution power-law 
exponent at criticality.

\subsection{Crossover behavior near criticality}

\begin{center}\includegraphics[%
  width=2.5in,
  height=2.0in]{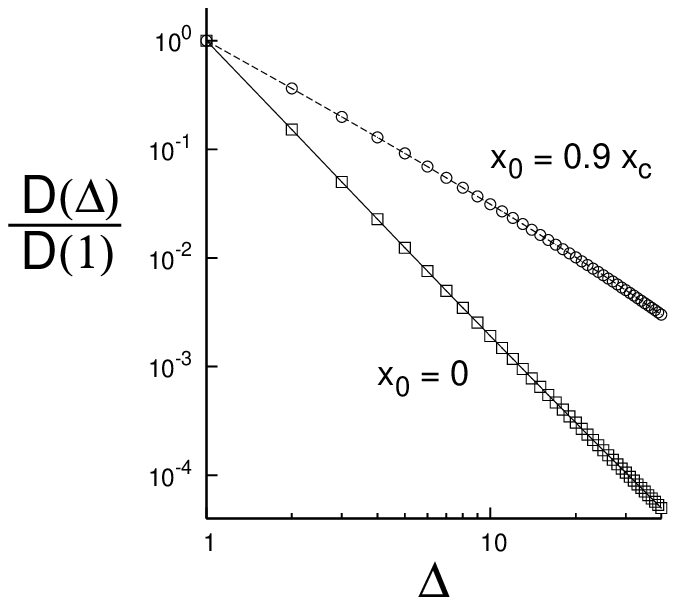}\end{center}

\noindent {\small Fig.\ 7. The distribution of bursts for threshold's
uniformly distributed in an interval $(x_{0},x_{c})$, with $x_{0}=0$
and with $x_{0}=0.9x_{c}$. The figure is based on 50 000 samples,
each with $N=10^{6}$ fibers.}\\

When \emph{all} fiber failures are recorded we have seen that the
burst distribution $D(\Delta)$ follows the asymptotic power law $D\propto\Delta^{-5/2}$.
If we just sample bursts that occur near criticality, a different behavior
is seen \cite{HHP-Pradhan1,HHP-Pradhan2} . As an illustration we consider the uniform threshold distribution,
and compare the complete burst distribution with what one gets when
one samples merely bursts from breaking fibers in the threshold interval
$(0.9x_{c},x_{c})$. Fig.\ 7 shows clearly that in the latter case
a different power law is seen.\\

If we want to study specifically the contribution from failures
occurring when the bundle is nearly critical, we evaluate the expression
(\ref{HHP-saddle}) for the burst distribution over a small interval $(x_{0},x_{c})$,
rather than integrating from $0$ to $x_{c}$. The argument in Sec.\ 2.1
that the major contribution to the integral comes from the critical
neighborhood is still valid. We obtain 
\begin{equation}
\frac{D(\Delta)}{N}=\frac{\Delta^{\Delta-2}e^{-\Delta}}{\Delta!}\;\frac{p(x_{c})}{a'(x_{c})}\left[1-e^{-\Delta/\Delta_{c}}\right],\label{HHP-D3}\end{equation}
 with \begin{equation}
\Delta_{c}=\frac{2}{a'(x_{c})^{2}(x_{c}-x_{0})^{2}}.\label{HHP-Dc}\end{equation}

By use of Stirling approximation $\Delta!\simeq\Delta^{\Delta}e^{-\Delta}\sqrt{2\pi\Delta}$,
-- a reasonable approximation even for small $\Delta$ -- the burst
distribution (\ref{HHP-D3}) may be written \begin{equation}
\frac{D(\Delta)}{N}=C\Delta^{-5/2}\left(1-e^{-\Delta/\Delta_{c}}\right),\label{HHP-D2}\end{equation}
 with a nonzero constant \begin{equation}
C=(2\pi)^{-1/2}p(x_{c})/a'(x_{c}).\end{equation}
 We see from (\ref{HHP-D2}) that there is a crossover at a burst length
around $\Delta_{c}$, so that \begin{equation}
\frac{D(\Delta)}{N}\propto\left\{ \begin{array}{cl}
\Delta^{-3/2} & \mbox{ for }\Delta\ll\Delta_{c}\\
\Delta^{-5/2} & \mbox{ for }\Delta\gg\Delta_{c}\end{array}\right.\end{equation}
The difference between the two power-law exponents is unity, as suggested by 
Sornette's ``sweeping of an instability"  mechanism \cite {HHP-sornette-sweep}.
Such a difference in avalanche power law exponents has  been observed 
numerically by Zapperi et al. in a fuse model \cite{HHP-Zapperi-crossover}.
 
We have thus shown the existence of a crossover\index{crossover} from the 
generic asymptotic behavior $D\propto\Delta^{-5/2}$ to the power law 
$D\propto\Delta^{-3/2}$
near criticality, i.e., near global breakdown. The crossover is a
universal phenomenon, independent of the threshold distribution $p(x)$.
In addition we have located where the crossover takes place. 

For the uniform distribution\index{uniform distribution} $\Delta_{c}=(1-x_{0}/x_{c})^{-2}/2$,
so for $x_{0}=0.8\, x_{\textrm{c}}$, we have $\Delta_{c}=12.5$.
For the Weibull distribution\index{Weibull distribution} $P(x)=1-\exp(-(x-1)^{10})$, where $1\leq x\leq\infty$,
we get $x_{c}=1.72858$ and for $x_{0}=1.7$, the crossover point
will be at $\Delta_{c}\simeq14.6$. Such crossover is clearly observed
(Fig. 8) near the expected values $\Delta=\Delta_{c}=12.5$ and $\Delta=\Delta_{c}=14.6$,
respectively, for the above distributions. 

\begin{center}\includegraphics[%
  width=2.0in,
  height=1.9in]{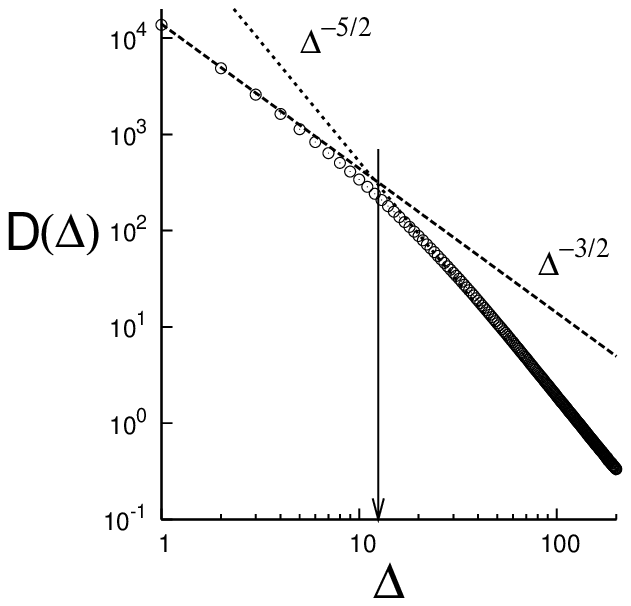}\hskip.2in\includegraphics[%
  width=2.0in,
  height=1.9in]{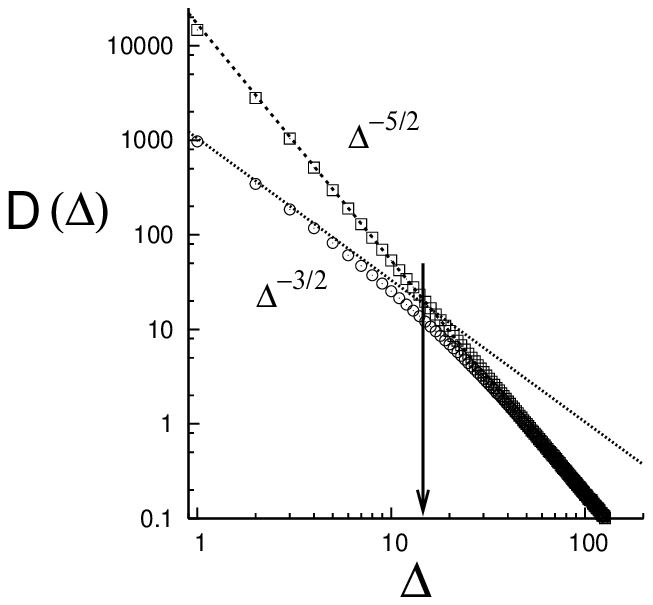}\end{center}

\noindent {\small Fig.\ 8. The distribution of bursts for the uniform
threshold distribution (left) with $x_{0}=0.80x_{\textrm{c}}$ and
for a Weibull distribution (right) with $x_{0}=1$ (square) and $x_{0}=1.7$
(circle). Both the figures are based on $50000$ samples with $N=10^{6}$
fibers each. The straight lines represent two different power laws,
and the arrows locate the crossover points $\Delta_{c}\simeq12.5$}
{\small and $\Delta_{c}\simeq14.6$,} respectively.\\

The simulation results shown in the figures are based on \textit{averaging}
over a large number of fiber bundles with moderate $N$. For applications
it is important that crossover\index{crossover} signals are seen also in a single sample.
We show in Fig.\ 9 that equally clear power laws are seen in a \textit{single}
fiber bundle when $N$ is large. 

\begin{center}\includegraphics[%
  width=2.2in,
  height=2.0in]{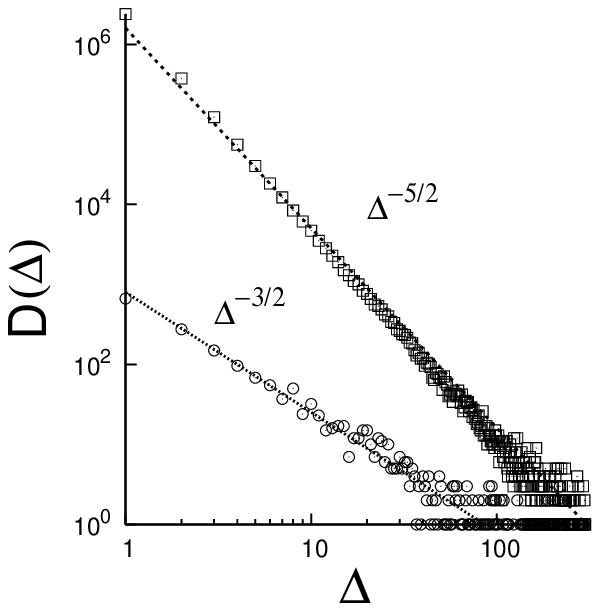}\end{center}

\noindent {\small Fig.\ 9. The distribution of bursts for the uniform 
threshold distribution for a single fiber bundle with $10^{7}$ fibers. 
Results with $x_{0}=0$  (recording all avalanches), are shown as squares, the 
circles stand for avalanches near the critical point ($x_{0}=0.9x_{c}$). }
{\small}\\

An important question in strength considerations of materials is
how to obtain signatures that can warn of imminent system failure\index{imminent failure}.
This is of uttermost importance in, {\em e.g.}, the dimond mining
industry where sudden failure of the mine can be very costly in terms
of lives. These mines are under continuous acoustic surveillance, but
at present there are no tell-tale acoustic signature of imminent
catastrophic failure. The same type of question is of course also
central to earthquake\index{earthquake} prediction. The crossover seen here in our fiber
bundle models is such a signature, it signals that catastrophic failure
is imminent. The same type of crossover phenomenon is also seen in
the burst distribution of a two-dimensional model of fuses\index{fuse model} 
with
stochastically distributed current tresholds \cite{HHP-Pradhan1}. This
signal does not hinge on observing rare events, and is seen also in
a single system (Fig.\ 9). It has, therefore, a strong potential as a
useful detection tool\index{detection tool}. 
It is interesting that most recently, Kawamura 
\cite{HHP-Kawamura} has observed a decrease in exponent 
value of the local magnitude distribution of earthquakes as the 
mainshock\index{mainshock} is 
approached  (See Fig. 20 of Ref.\cite{HHP-Kawamura}), analysing earthquakes 
in Japan (from JUNEC catalog). 

   Obviously, one cannot count bursts all the way to complete
 breakdown to have a useful detection tool. It suffices to sample
 bursts in  finite intervals $(x_0,x_f)$,
 with $x_f<x_c$. In this case we obtain the avalanche distribution by
 restricting the integration in Eq.(18) to the appropiate intervals.
 When such an interval is in the neighborhood of $x_c$ we obtain
 \begin{eqnarray}
 \frac{D(\Delta)}{N} &\simeq &
 \frac{\Delta^{\Delta-1}e^{-\Delta}p(x_c)a'(x_c)}{\Delta !}
 \int_{x_0}^{x_f}e^{- a'(x_c)^2(x_c-x)^2\Delta/2}\;(x-x_c)\;dx\\
   &\propto & \Delta^{-5/2} \left(e^{- \Delta
 (x_c-x_f)^2/a}-e^{-\Delta (x_c-x_0)/a}\right),
 \end{eqnarray}
 with $a=2/a'(x_c)^2$.
 This shows a crossover:
 \begin{equation}
 \frac{D(\Delta)}{N} \propto \left\{ \begin{array}{ll}
  \; \Delta^{-3/2}& \mbox{ for } \Delta \ll a/(x_c-x_0)^2 \\
  \Delta^{-5/2} & \mbox{ for } a/(x_c-x_0)^2 \ll \Delta \ll
 a/(x_c-x_f)^2,
 \end{array} \right.
 \end{equation}
 with a final exponential behavior when $\Delta \gg a/(x_c-x_f)^2$.

 The 3/2 power law will be seen only when the beginning of the
 interval, $x_0$, is close enough to the critical value $x_c$.
 Observing the $3/2$ power law is therefore a signal of imminent
 system failure.

\subsection{Avalanche distribution at criticality}

Precisely \emph{at} criticality $(x_{0}=x_{c})$ the crossover takes
place at $\Delta_{c}=\infty$, and consequently the $\xi=5/2$ power
law is no longer present. We will now argue, using the random walk
representation in section 2.3, that precisely at criticality the avalanche
distribution follows a power law with exponent $3/2$. 

At criticality the distribution (\ref{HHP-step}) of the step lengths
in the random walk simplifies to \begin{equation}
\rho_{c}(\Delta F)=\left\{ \begin{array}{cl}
0 & \mbox{ for }\Delta F<-x_{c}\\
x_{c}^{-1}e^{-1}e^{-\Delta F/x_{c}} & \mbox{ for }\Delta F\geq-x_{c}\end{array}\right.\label{HHP-criticalwalk}\end{equation}

A first burst of size $\Delta$ corresponds to a random walk\index{random walk} in which
the position after each of the first $\Delta-1$ steps is \emph{lower}
than the starting point, but after step no. $\Delta$ the position
of the walker exceeds the starting point. The probability of this
equals \begin{eqnarray}
\mbox{Prob}(\Delta) & = & \int_{-x_{c}}^{0}\rho_{c}(\delta_{1})d\delta_{1}\int_{x_{c}}^{-\delta_{1}}\rho_{c}(\delta_{2})d\delta_{2}\int_{-x_{c}}^{-\delta_{1}-\delta_{2}}\rho_{c}(\delta_{3})d\delta_{3}\ldots\nonumber \\
 &  & \int_{-x_{c}}^{-\delta_{1}-\delta_{2}\ldots-\delta_{\Delta-2}}\rho_{c}(\delta_{\Delta-1})d\delta_{\Delta-1}\int_{-\delta_{1}-\delta_{2}\ldots-\delta_{\Delta-1}}^{\infty}\rho_{c}(\delta_{\Delta})d\delta_{\Delta}.\label{HHP-criticalRW}\end{eqnarray}
 To simplify the notation we have introduced $\delta_{n}\equiv\Delta F_{n}$.
In Ref. \cite{HHP-Pradhan2} we have evaluated the multiple integral (\ref{HHP-criticalRW}),
with the result \begin{equation}
\mbox{Prob}(\Delta)=\frac{\Delta^{\Delta-1}\, e^{-\Delta}}{\Delta!}\simeq\frac{1}{\sqrt{2\pi}}\;\Delta^{-3/2}.\label{HHP-RWc}\end{equation}

We note in passing that for the standard unbiased random walk with
constant step length we obtain a \emph{different} expression for the
burst probability, but again with a 3/2 power law for large $\Delta$:
\begin{equation}
\mbox{Prob}(\Delta)=\frac{1}{2^{\Delta-1}\,\Delta}\left(\begin{array}{c}
\Delta-2\\
\frac{1}{2}\Delta-1\end{array}\right)\simeq\frac{1}{\sqrt{2\pi}}\;\Delta^{-3/2}.\end{equation}

At completion of the first burst, the force, i.e., the excursion of
the random walk, is larger than all previous values. Therefore one
may use this point as a new starting point to find, by the same calculation,
the distribution of the next burst, etc. Consequently the complete
burst distribution is essentially proportional to $\Delta^{-3/2}$,
as expected. The simulation results exhibited in Fig.\ 10 are in excellent
agreement with these predictions.

\begin{center}\includegraphics[%
  width=2.2in,
  height=2.0in]{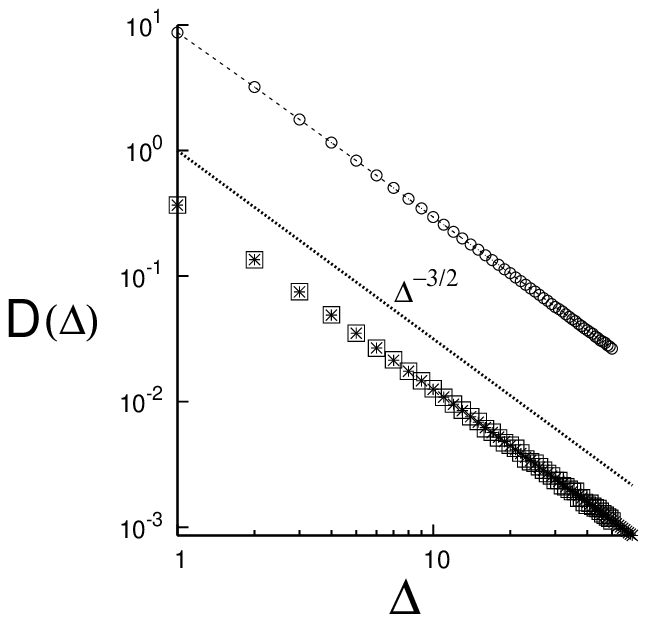}\end{center}

\noindent {\small Fig.\ 10. Distributions of the first bursts (squares)
and of all bursts (circles) for the uniform threshold distribution
with $x_{0}=x_{c}$. The simulation results are based on $10^{6}$
samples with $80000$ fibers each. The crosses stand for the analytic
result (\ref{HHP-RWc}).}\\

One of the unusual threshold distributions\index{threshold distribution} we studied in Sec.\ 2.2
corresponded to an constant average force\index{average force}, $\langle F\rangle/N$ independent
of $x$. Such a bundle is not critical at a single point $x_{c}$,
but in a whole interval of $x$. That the burst exponent for this
model takes the critical value $3/2$ is therefore no surprise.

\subsection{Recursive dynamics}

The relation between the number of ruptured fibers and a given external
load per fiber, $\sigma=F/N$, can be viewed as the result of a sequential
process. In the first step all fibers with thresholds less than $x_{1}=\sigma$
must fail. Then the load is redistributed on the surviving fibers,
which forces more fibers to burst, etc. This starts an iterative process
that goes on until equilibrium is reached, or all fibers rupture\index{rupture} \cite{HHP-Pradhan4,HHP-Pradhan5,HHP-Pradhan6}. 

Assume all fibers with thresholds less than $x_{t}$ break in step
number $t$. The expected number of intact fibers is then \begin{equation}
U_{t}=1-P(x_{t}),\label{HHP-rec-1}\end{equation}
 so that the load per fiber is increased to $\sigma/U_{t}$. In step
number $t+1$, therefore, all fibers with threshold less than \begin{equation}
x_{t+1}=\frac{\sigma}{1-P(x_{t})}\label{HHP-iteration}\end{equation}
 must fail. This iteration defines the recursive dynamics\index{recursive dynamics} \cite{HHP-Pradhan4,HHP-Pradhan5,HHP-Pradhan6}.
Alternatively an iteration for the $U_{t}$ can be set up: \begin{equation}
U_{t+1}=1-P(\sigma/U_{t});\hspace{5mm}U_{0}=1.\label{HHP-Uiteration}\end{equation}

If the iteration (\ref{HHP-iteration}) converges to a finite fixed-point\index{fixed-point}
$x^{*}$, \[
\lim_{t\rightarrow\infty}x_{t}=x^{*},\]
 the fixed-point value must satisfy \begin{equation}
x^{*}=\frac{\sigma}{1-P(x^{*})}.\end{equation}
 This fixed-point relation, \begin{equation}
\sigma=x^{*}\,[1-P(x^{*})],\label{HHP-fixedpoint}\end{equation}
 is identical to the relation (\ref{HHP-4}) between the average force
per fiber, $\langle F\rangle/N$, and the threshold value $x$. 

Eq.\ (\ref{HHP-fixedpoint}) shows that a necessary condition to have
a finite positive fixed-point value $x^{*}$ is \begin{equation}
\sigma\leq\sigma_{c}\equiv\max_{x}\;\left\{ x\,[1-P(x)]\right\} \end{equation}
 Thus $\sigma_{c}$ is the critical value of the external load per
fiber, beyond which the bundle fails completely. 

As a simple example take the uniform threshold distribution\index{uniform distribution} (\ref{HHP-uniform}) with $x_{r}=1$,
for which the iterations take the form \begin{equation}
x_{t+1}=\frac{\sigma}{1-x_{t}}\hspace{5mm}\mbox{ and }U_{t+1}=1-\frac{\sigma}{U_{t}}.\label{HHP-iterationuni}\end{equation}
 Moreover, $\sigma_{c}=1/4$, and the quadratic fixed-point equation
for $U^{*}$ has the solution \begin{equation}
U^{*}(\sigma)={\textstyle \frac{1}{2}}\pm(\sigma_{c}-\sigma)^{1/2}=U^{*}(\sigma_{c})\pm(\sigma_{c}-\sigma)^{1/2},\label{HHP-Ufixed}\end{equation}
 since $U^{*}(\sigma_{c})=1/2$. The iterations are sketched in Fig.\ 11. 

\begin{center}\includegraphics[%
  width=2.0in,
  height=2.2in]{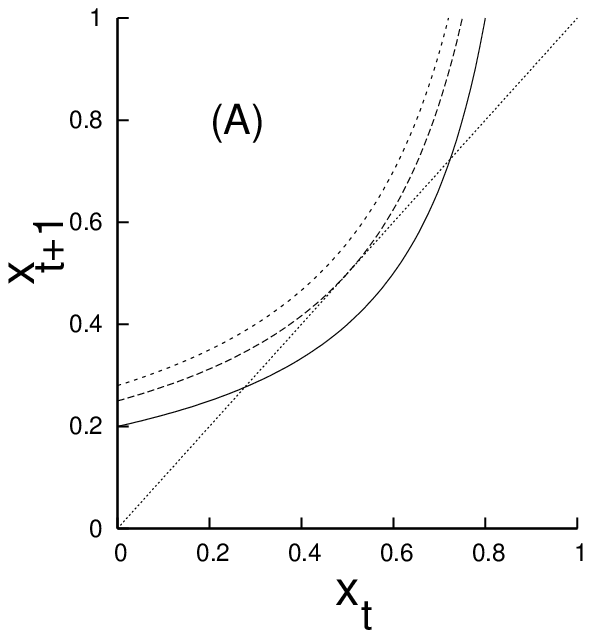}\hskip.2in\includegraphics[%
  width=2.0in,
  height=2.2in]{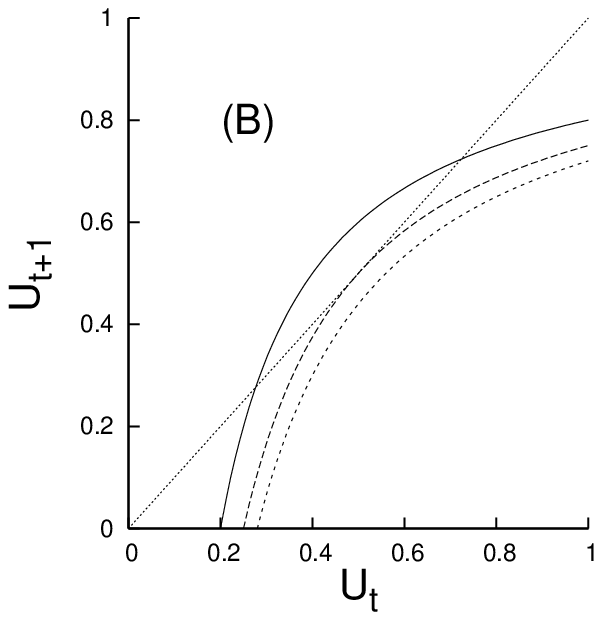}\end{center}

\noindent {\small Fig.\ 11. Graphical representation of the iterations
for $x$ (A) and for $U$ (B). The graphs are shown for different
values of the external stress: $\sigma=0.2$ (solid), $0.25$ (dashed),
and $0.28$ (dotted), respectively. The intersections between the
graph and the straight $45^{\circ}$ line define possible fixed points.}\\

A fixed point is attractive if $|dU_{t+1}/dU_{t}|$ is less than $1$
at the fixed point and repulsive if the derivative exceeds unity.
We see in Fig.\ 11 that the stable fixed point\index{fixed-point} corresponds to the
smallest value of $x^{*}$, and to the largest value of $U^{*}$ (the
plus sign in (\ref{HHP-Ufixed})): \begin{equation}
U^{*}(\sigma)-U^{*}(\sigma_{c})=(\sigma_{c}-\sigma)^{\beta},\hspace{5mm}\mbox{with }\hspace{3mm}\beta={\textstyle \frac{1}{2}}.\end{equation}
 Thus $U^{*}(\sigma)-U^{*}(\sigma_{c})$ behaves like an order parameter\index{order parameter},
signalling total bundle failure when it is negative, partial failure
when it is positive. 

Close to a stable fixed point the iterated quantity changes with tiny
amounts, so one may expand in the difference $\epsilon_{t}=U_{t}-U^{*}$.
To first order (\ref{HHP-iterationuni}) yields \begin{equation}
\epsilon_{t+1}=\epsilon_{t}\cdot\frac{\sigma}{U^{*2}}.\end{equation}
 Thus the fixed point is approached monotonously, with exponentially
decreasing steps: \begin{equation}
\epsilon_{t}\propto e^{-t/\tau},\end{equation}
 with \begin{equation}
\tau=\frac{1}{\ln(U^{*2}/\sigma)}.\end{equation}
 Precisely \emph{at} the critical point, where $U^{*}=1/2$ and $\sigma_{c}=1/4$,
the relaxation parameter $\tau$ is infinite, signalling a non-exponential
approach to the fixed point. Close to the critical point one easily
shows that \begin{equation}
\tau\propto(\sigma_{c}-\sigma)^{-\alpha},\hspace{5mm}\mbox{with}\hspace{3mm}\alpha={\textstyle \frac{1}{2}}.\end{equation}

One may define a \emph{breakdown susceptibility} $\chi$ by the change
of $U^{*}(\sigma)$ due to an infinitesimal increment of the applied
stress $\sigma$, \begin{equation}
\chi=-\frac{dU^{*}(\sigma)}{d\sigma}={\textstyle \frac{1}{2}}(\sigma_{c}-\sigma)^{-\gamma},\hspace{5mm}\mbox{with}\hspace{3mm}\gamma={\textstyle \frac{1}{2}}.\end{equation}
 The susceptibility\index{susceptibility} diverges as the applied stress $\sigma$ approaches
its critical\index{critical point} value. Such a divergence was noted in previous numerical
studies \cite{HHP-Zapperi,HHP-Silveira}. 

When at criticality the approach to the fixed point is not exponential,
what is it? Putting $U_{t}=U_{c}+\epsilon_{t}$ in the iteration (\ref{HHP-iterationuni})
for $\sigma=1/4$, it may be rewritten as follows \begin{equation}
\epsilon_{t+1}^{-1}=\epsilon_{t}^{-1}+2,\hspace{5mm}\mbox{with}\hspace{3mm}\epsilon_{0}={\textstyle \frac{1}{2}},\end{equation}
 with solution $\epsilon_{t}^{-1}=2t+2$. Thus we have, \emph{exactly},
\begin{equation}
U_{t}=\frac{1}{2}+\frac{1}{2t+2}.\end{equation}
 For large $t$ this follows a power-law approach to the fixed point,
$U_{t}-U_{c}=\frac{1}{2}t^{-\delta}$, with $\delta=1$. 

These critical properties are valid for the uniform distribution,
and the natural question is how general the results are. In Ref. \cite{HHP-Pradhan6}
two other threshold distributions were investigated, and all critical
properties, quantified by the indicies $\alpha, \beta, \gamma$ and  $\delta$
were found to be the same as for the uniform threshold distribution. This
suggests strongly that the critical behavior is universal\index{universal}, which we now prove. 

When an iteration is close to the fixed point, we have for the deviation
\begin{equation}
\epsilon_{t+1}=U_{t+1}-U^{*}=P\left(\frac{\sigma}{U^{*}}\right)-P\left(\frac{\sigma}{U^{*}+\epsilon_{t}}\right)=\epsilon_{t}\cdot\frac{\sigma}{U^{*2}}p(\sigma/U^{*}),\label{HHP-dev-1}\end{equation}
to lowest order in $\epsilon_{t}$.

This guarantees an exponential relaxation\index{relaxation} to the fixed point, $\epsilon_{t}\propto e^{-t/\tau}$,
with parameter \begin{equation}
\tau=1\left/\ln\left(\frac{U^{*2}}{\sigma p(\sigma/U^{*})}\right)\right..\end{equation}
 Criticality is determined by the extremum condition (8), which by
the relation (\ref{HHP-rec-1}) takes the form \[
U_{c}^{2}=\sigma p(\sigma/U_{c})\]
 Thus $\tau=\infty$ at criticality. To study the relaxation at criticality
we must expand (\ref{HHP-dev-1}) to second order in $\epsilon_{t}$ since
to first order we simply get the useless equation $\epsilon_{t+1}=\epsilon_{t}$.
To second order we obtain \[
\epsilon_{t+1}=\epsilon_{t}-C\epsilon_{t}^{2},\]
 with a positive constant $C$. This is satisfied by \[
\epsilon_{t}=\frac{1}{Ct}+\mathcal{O}(t^{-2}).\]
 Hence in general the dominating critical behavior for the approach
to the fixed point is a power law with $\delta=1$. The values $\alpha=\beta=\gamma=\frac{1}{2}$
can be shown to be consequences of the parabolic maximum of the load
curve, (7) or (49), at criticality: $\left|x_{c}-x^{*}\right|\propto(\sigma_{c}-\sigma)^{\frac{1}{2}}$. 

Thus all threshold distributions for which the macroscopic strength
function has a single parabolic maximum, are in this universality class\index{universality class}.

\section{Fiber bundles with local load redistribution}

The assumption that the extra stress caused by a fiber failure is
shared equally among all surviving fibers is often unrealistic, since
fibers in the neighborhood of the failed fiber are expected to take
most of the load increase. One can envisage many systems for such
local stress redistributions. A special case is the model with a one-dimensional
geometry where the two nearest-neighbor fibers take up all extra stress
caused by a fiber failure (Sec.\ 3.1). It is special for two reasons:
It is an extreme case because the range of the stress redistribution
is minimal, and, secondly, it is amenable to theoretical analysis.
In other models, treated in Sec. 3.2 and 3.3, the stress redistribution
occurs over a larger region. In Sec. 3.3 this comes about by considering
a clamp to be an elastic medium.

\subsection{Stress alleviation by nearest neighbors}

The simplest model with nearest-neighbor stress 
redistribution is one-dimensional,
with the $N$ fibers ordered linearly, with or without periodic boundary
conditions. Thus two fibers, one of each side, take up, and divide
equally, the extra stress caused by a failure. The force on a fiber
surrounded by $n_{l}$ broken fibers on the left-hand side and $n_{r}$
broken fibers on the right-hand side is then \begin{equation}
\frac{F_{tot}}{N}\left(1+{\textstyle \frac{1}{2}}n_{l}+{\textstyle \frac{1}{2}}n_{r}\right)\equiv f(2+n_{l}+n_{r}),\label{HHP-neighbor}\end{equation}
 where $F_{tot}$ is the total force on the bundle, and $f=F_{tot}/2N$,
one-half the force-per-fiber, is a convenient forcing parameter. The
model has been discussed previously in a different context\cite{HHP-Harlow,HHP-Harlow2,HHP-Duxbury,HHP-Harlow3,HHP-Phoenix,HHP-Kuo}.
Preliminary simulation studies \cite{HHP-Hansen,HHP-Ding} showed convincingly
that this local model is \emph{not} in the same universality class
as the equal-load-sharing fiber bundles. For the uniform threshold
distribution and for $1\leq\Delta\leq10$ an effective exponent between
4 and 5 was seen, much larger than $5/2$. 

Avalanches\index{avalanche} in this model, and in similar local stress-redistribution
models, have a character different from bursts in the equal-load-sharing
models. In the present model the failure of one fiber can by a domino effect\index{domino effect}
set in motion a fatal avalanche: If the failing fiber has many previously
failed fibers as neighbors, the load on the fibers on each side is
high, and if they burst, the load on the new neighbors is even higher,
etc., which may produce an unstoppable avalanche. 

In Ref.\cite{HHP-Kloster} the burst distribution\index{burst distribution} was determined analytically
for the uniform threshold distribution\index{uniform distribution}, \begin{equation}
P(x)=\left\{ \begin{array}{cl}
x & \mbox{ for }0\leq x\leq1\\
1 & \mbox{ for }x>1\end{array}\right.\end{equation}

In this model there is an upper limit to the size $\Delta$ of an
avalanche that the bundle can survive. Since the threshold values
are at most unity, it follows from (\ref{HHP-neighbor}) that if \begin{equation}
\Delta>f^{-1}-2,\end{equation}
 then the bundle breaks down completely. Consequently an asymptotic
power law distribution of the avalanche sizes is not possible. 

In the analytic derivation periodic boundary conditions were used.
The fairly elaborate procedure was based on a set of recursion relations
between configurations and events at fixed external force. In Fig.\ 12
we show the resulting burst distribution for a bundle of $N=20000$
fibers, compared with simulation results. The agreement is extremely
satisfactory.

\vskip-.2in

\begin{center}\includegraphics[%
  width=3.0in,
  height=3.0in,
  angle=-90]{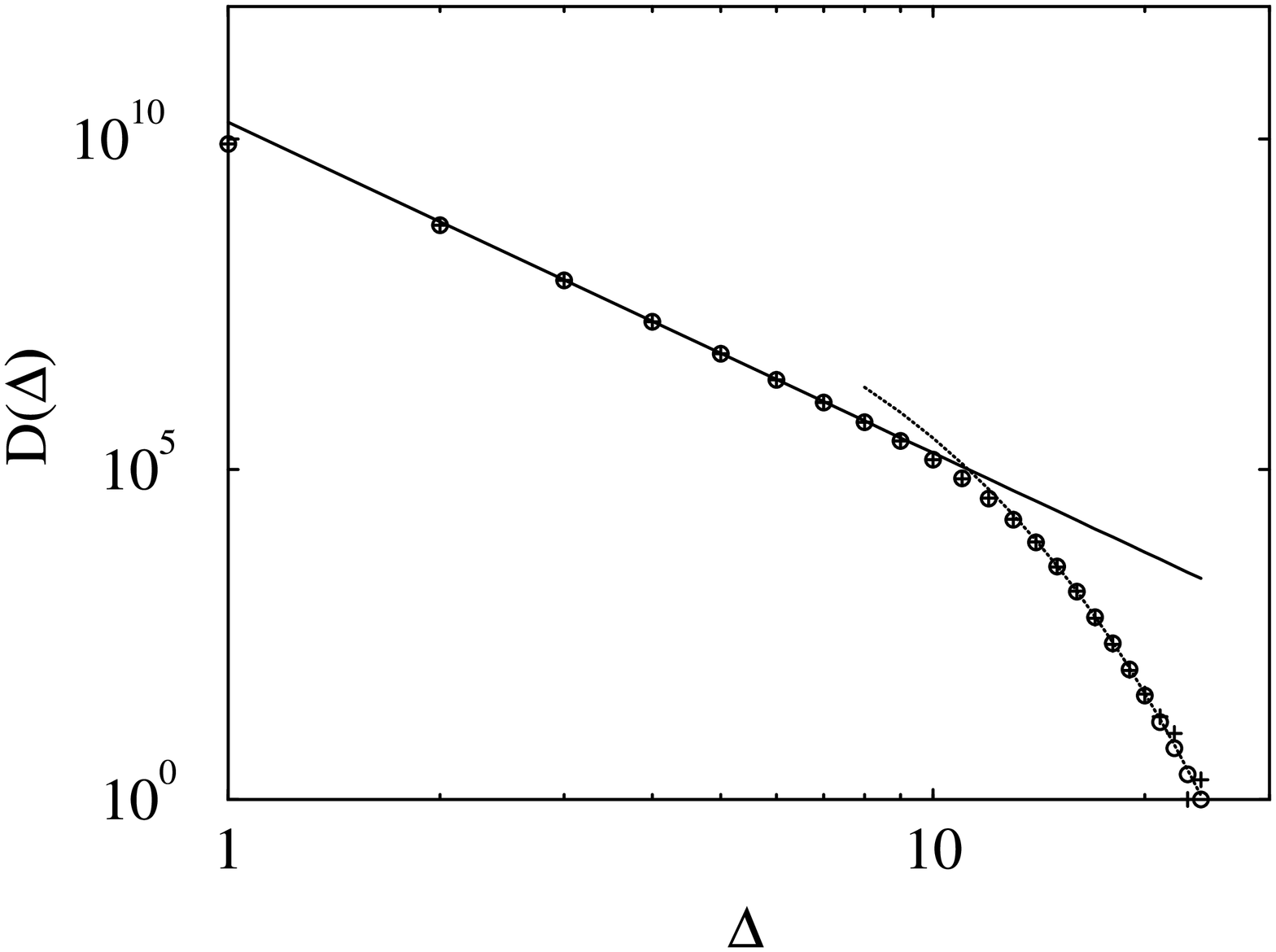}\end{center}

{\small Fig.\ 12. Burst distribution in a local-load-sharing model
for a bundle of ${N=20000}$ fibers. Theoretical results are shown as
circles, and simulation results (for ${4 000 000}$ samples) are shown
as crosses. The straight line shows the power law $\Delta^{-5}$.}\\

We see from Fig. 12 that, as expected, the burst distribution does
\emph{not} follow a power law for large $\Delta$, but falls off faster.
For $\Delta$ less than 10 the burst distribution follows approximately
a power law\index{power-law}, $D(\Delta)\propto\Delta^{-\xi}$ with $\xi$ of the order of 5. 

In Ref.\cite{HHP-Kloster} the maximum load $F_{max}$ the fiber bundle
could tolerate was estimated to have the following size dependence,
\begin{equation}
F_{max}\propto\frac{N}{\ln N}.\end{equation}
 This is different from the equal-load-sharing model, for which $F_{max}\propto N$.
A similar logarithmic size-dependence of the bundle strength for local-stress-redistribution
models has been proposed by other authors \cite{HHP-Smith,HHP-Zhang,HHP-Zhang2}. 

The qualitative explanation of the non-extensive result is that for
large $N$ the probability of finding a weak region somewhere is high.
Since, as discussed above, a weak region is the seed to complete bundle
failure, it may be reasonable that the maximum load the bundle can
carry does not increase proportional to $N$, but slower than linear.

\subsection{Intermediate load-sharing models}

It might be interesting to study models that interpolates between
global and nearest-neighbor load sharing\index{intermediate load sharing}. The main question is whether
the burst distribution changes from one behavior to the other in a
continuous manner, or whether a discontinuous change occurs. 

In Ref.\cite{HHP-Hansen2} was introduced such an intermediate model,
with the same one dimensional geometry as the nearest-neighbor model
of the preceding section. When a fiber $i$ fails in this model the
elastic constants of the two nearest surviving neighbors $l$ and
$r$ on both sides are updated as follows \begin{eqnarray}
\kappa_{l} & \rightarrow & {\textstyle \frac{1}{2}}\lambda(\kappa_{l}+\kappa_{r}+\kappa_{i})\\
\kappa_{i} & \rightarrow & 0\\
\kappa_{r} & \rightarrow & {\textstyle \frac{1}{2}}\lambda(\kappa_{l}+\kappa_{r}+\kappa_{i}).\end{eqnarray}
 For $\lambda=1$ this corresponds to the local load-sharing by surviving
nearest-neighbors (see the preceding section). And with $\lambda=0$
the intact neares- neighbor fibers to a failing fiber does not take
part in the load-sharing. But since all the other surviving fibers
then share the load equally, this limiting case must have the same
behavior as the equal-load-sharing model. The numerics seems to suggest
that there is a cross-over value of $\lambda$ separating the universality
classes\index{universality class} of the local-load-sharing and the equal-load-sharing regimes. 

A stress redistribution scheme that in a straightforward way interpolates
between the two extreme models was recently proposed by Pradhan et
al.\cite{HHP-Pradhan3}: A fraction $g$ of the extra load caused by a
fiber failure is shared by the nearest neighbors, and the remaining
load is shared equally among all intact fibers. They show that in
a one-dimensional geometry a crossover value $g_{c}$ exists, such
that for $g<g_{c}$ the bundle belongs to the equal-load-sharing regime,
while for $g>g_{c}$ the system is like the local-load-sharing model
of Sec.\ 3.1. The crossover\index{crossover} value was determined to be $g_{c}=0.79\pm0.01$. 

It would be more realistic to have a stress redistribution whose magnitude
falls off monotonically with the geometric distance $r$ from the
failed fiber. Hidalgo et al. \cite{HHP-Hidalgo} introduced such a model,
for which the extra stress on a fiber followed a power law decay,
proportional to $r^{-\gamma}$. In the limit $\gamma\rightarrow0$
the equal-load-distribution model is recovered, while the limit $\gamma\rightarrow\infty$
corresponds to the nearest-neighbor model in Sec.\ 3.1. Again a crossover
is observed, at a value $\gamma_{c}\simeq2$ for the range parameter. 

In the next section we consider a model with a different, but similar,
interaction decaying with the distance from the failed fiber.

\subsection{Elastic medium anchoring}
In this section we generalize the fiber bundle problem to include more
realistically  the elastic response\index{elastic response} of the surfaces to which the fibers are
attached.  So far, these have been assumed to be infinitely stiff for
the equal-load-sharing model, or their response has been modeled  as very
soft, but in a fairly unrealistic way in the local-load-sharing models, see
Section 3.2.  In \cite{HHP-bhs02}, a realistic model for the elastic response
of the clamps was studied.  The model was presented as addressing the problem
of failure of weldings. In this context, the two clamps were seen as
elastic media glued together at a common interface. Without loss of generality,
one of the media was assumed to be infinitely stiff whereas the other
was soft.

The two clamps can be pulled apart by controlling (fixing) either the
applied force or the {\it displacement}. The displacement is defined
as the change in the distance between two points, one in each clamp
positioned far from the interface. The line connecting
these points is perpendicular to the average position of the
interface. In our case, the pulling is accomplished by controlling the
displacement. As the displacement is increased slowly, fibers ---
representing the glue --- will fail, ripping the two surfaces apart.

The model consists of two
two-dimensional square $L\times L$ lattices with periodic boundary
conditions. The lower one represents the hard, stiff surface and the
upper one the elastic surface. The nodes of the two lattices are
matched ({\it i.e.\/} there is no relative lateral displacement). The
thresholds of the fibers are taken from an uncorrelated uniform distribution.
The spacing between the fibers
is $a$ in both the $x$ and $y$ directions.  The force that each fiber
is carrying is transferred over an area of size $a^2$ to the soft clamp:
As the two clamps are
separated by controlling the displacement of the hard clamp, $D$,
the forces carried by the fibers increase. As for the fiber bundle models
studied in the previous sections, when the force
carried by a fiber reaches its breaking threshold, it breaks
irreversibly and the forces redistribute.  Hence, the fibers are broken
one by one until the two clamps are no longer in mechanical contact. As
this process is proceeding, the elastic clamp is of course deforming to
accomodate the changes in the forces acting on it.

The equations governing the system are as follows. The force, $f_i$,
carried by the $i$th fiber is given by
\begin{equation}
\label{HHP-M1}
f_i = -k(u_i-D)\;,
\end{equation}
where $k$ is the spring constant
and $u_i$ is the deformation of the elastic clamp at site
$i$. All unbroken fibers have $k=1$ while a broken fiber has
$k=0$. The quantity $(u_i-D)$ is, therefore, the length fiber $i$ is
stretched. In addition, a force applied at a point on an elastic
surface will deform this surface over a region whose extent depends on
its elastic properties. This is described by the coupled system of
equations,
\begin{equation}
\label{HHP-M2}
u_i=\sum_{j} G_{i,j} f_j\;,
\end{equation}
where the elastic Green function, $G_{i,j}$ is given by \cite{HHP-ll58,HHP-j85}
\begin{equation}
\label{HHP-M3}
G_{i,j} = \frac{1-s^2}{\pi e a^2}\ \int_{-a/2}^{+a/2}\int_{-a/2}^{+a/2}\
\frac{dx'\ dy'}{|(x-x',y-y')|}\;.
\end{equation}
In this equation, $s$ is the Poisson ratio, $e$ the elastic constant,
and the denominator $|{\vec i}-{\vec j}|$ is the distance between sites $i=(x,y)$ and $j=(x',y')$. The
indices $i$ and $j$ run over all $L^2$ sites.  The integration over the
area $a^2$ is done to average the force from the fibers over this area.
Strictly speaking, the
Green function applies for a medium occupying the infinite half-space. 
However, with a judicious choice of elastic constants, we may
use it for a finite medium if its range is small compared to $L$, the
size of the system.

By combining equations (\ref{HHP-M1}) and (\ref{HHP-M2}), we obtain
\begin{equation}
\label{HHP-M4}
({\bf I} +{\bf K G}) {\vec f} = {\bf K} {\vec D}\;,
\end{equation}
where we are using matrix-vector notation. ${\bf I}$ is the $L^2\times
L^2$ identity matrix, and ${\bf G}$ is the Green function represented
as an $L^2\times L^2$ dense matrix. The constant vector ${\vec D}$ is
$L^2$ dimensional. The {\it diagonal\/} matrix ${\bf K}$ is also
$L^2\times L^2$. Its matrix elements are either 1, for unbroken
fibers, or 0 for broken ones. Of course, ${\bf K}$ and ${\bf G}$ do
not commute.

Once equation (\ref{HHP-M4}) is solved for the force ${\vec f}$, equation
(\ref{HHP-M2}) easily yields the deformations of the elastic clamp.

Equation (\ref{HHP-M4}) is of the familiar form ${\bf A}{\vec x}={\vec
b}$.  Since the Green function connects all nodes to all other nodes,
the $L^2\times L^2$ matrix ${\bf A}$ is dense which puts severe limits
on the size of the system that may be studied.

The simulation proceeds as follows: We start with all springs present,
each with its randomly drawn breakdown threshold. The two clamps are
then pulled apart, the forces calculated using the Conjugate
Gradient (CG) algorithm \cite{HHP-bh88,HHP-ptvf92}, and the fiber
which is the nearest to its threshold is broken, {\it i.e.\/} the  corresponding matrix element it the  matrix ${\bf K}$ is
zeroed. Then the new forces are calculated, a new fiber broken and so
on until all fibers have failed.

However, there are two problems that render the simulation of large
systems extremely difficult. The first is that since ${\bf G}$ is a
$L^2\times L^2$ {\it dense\/} matrix, the number of operations per CG
iteration scales like $L^4$. Even more serious is the fact that as the
system evolves and springs are broken, the matrix $({\bf I}+k{\bf G})$
becomes very ill-conditioned.

To overcome the problematic $L^4$ scaling of the algorithm, we note
that the Green function is diagonal in Fourier space. Consequently,
doing matrix-vector multiplications using FFTs the scaling is much
improved and goes like $L^2 \ln(L)$.  Symbolically, this can be
expressed as follow:
\begin{equation}
\label{HHP-M5}
({\bf I} + {\bf K} {\bf F^{-1}F}{\bf G}){\bf F^{-1}F} {\vec f} = {\bf
K} {\vec D}\;,
\end{equation}
where ${\bf F}$ is the FFT operator and ${\bf F^{-1}}$ its inverse
(${\bf F^{-1}F}=1$). Since ${\bf I}$ and ${\bf K}$ are diagonal,
operations involving them are performed in real space. With this
formulation, the number of operations/iteration in the CG algorithm
now scales like $L^2\ln(L)$.

To overcome the runaway behavior due to the ill-conditioning we need
to precondition the matrix \cite{HHP-bh88,HHP-bhn86}. This means that instead of
solving equation (\ref{HHP-M5}), we solve the equivalent problem
\begin{equation}
\label{HHP-M6}
{\bf Q}({\bf I} + {\bf K} {\bf F^{-1}F}{\bf G}){\bf F^{-1}F} {\vec f}
= {\bf Q} {\bf K} {\vec D}\;,
\end{equation}
where we simply have multiplied both sides by the arbitrary, positive
definite preconditioning matrix ${\bf Q}$. Clearly, the ideal choice
is ${\bf Q_{0}}=({\bf I} + {\bf K}{\bf G})^{-1}$ which would always
solve the problem in one iteration. Since this is not possible in
general, we look for a form for ${\bf Q}$ which satisfies the
following two conditions: (1) It should be as  close as possible to ${\bf Q_{0}}$,
and (2) be fast to calculate. The choice of a good ${\bf Q}$ is further
complicated by the fact that as the system evolves and fibers are
broken, corresponding matrix elements of ${\bf K}$ are set to
zero. So, the matrix $({\bf I} + {\bf K}{\bf G})$ evolves from the
initial form $({\bf I} + {\bf G})$ to the final one ${\bf I}$.

Batrouni et al.\ \cite{HHP-bhs02} chose the form
\begin{equation}
\label{HHP-M7}
{\bf Q}={\bf I}-({\bf K}{\bf G})+({\bf K}{\bf G})({\bf K}{\bf G})
-({\bf K}{\bf G})({\bf K}{\bf G})({\bf K}{\bf G})+ ...
\end{equation}
which is the Taylor series expansion of ${\bf Q_{0}}=({\bf
I} + {\bf K}{\bf G})^{-1}$. For best performance, the number of terms
kept in the expansion is left as a parameter since it depends on the
physical parameters of the system. It is important to emphasize the
following points: (a) As fibers are broken, the preconditioning
matrix evolves with the ill-conditioned matrix and, therefore, remains
a good approximation to its inverse throughout the breaking
process. (b) All matrix multiplications involving ${\bf G}$ are done
using FFTs. (c) The calculation of ${\bf Q}$ can be easily organized
so that it scales like $n L^2 \ln(L)$ where $n$ is the number of terms
kept in the Taylor expansion, equation (\ref{HHP-M7}).
The result is a stable accelerated algorithm which scales
essentially as the volume of the system.

Fig. 13 (left) shows the force-displacement curve for a system of
size $128\times 128$ with
elastic constant $e=10$.  Whether we control the applied force, $F$,
or the displacement, $D$, the system will eventually suffer
catastrophic collapse\index{catastrophic failure}.  However, this is not so when $e=100$ as shown
in Fig. 13 (right).  In this case, only controlling the force will lead
to catastrophic failure.  In the limit when $e\to\infty$, the model
becomes the equal-load-sharing fiber bundle model,
where $F=(1-D)D$.  In this limit there are no spatial correlations and
the force instability is due to the the decreasing total elastic
constant of the system making the force on each surviving bond
increase faster than the typical spread of threshold values. No such
effect exists when controlling displacement $D$.  However, when the
elastic constant, $e$, is small, spatial correlations in the form of
localization, where fibers that are close in space have a tendency to
fail consequtively, do develop, and these are responsible for the displacement
instability which is seen in Fig. 13.

\includegraphics[%
  width=2.5in,
  height=2.2in,
  angle=-90]{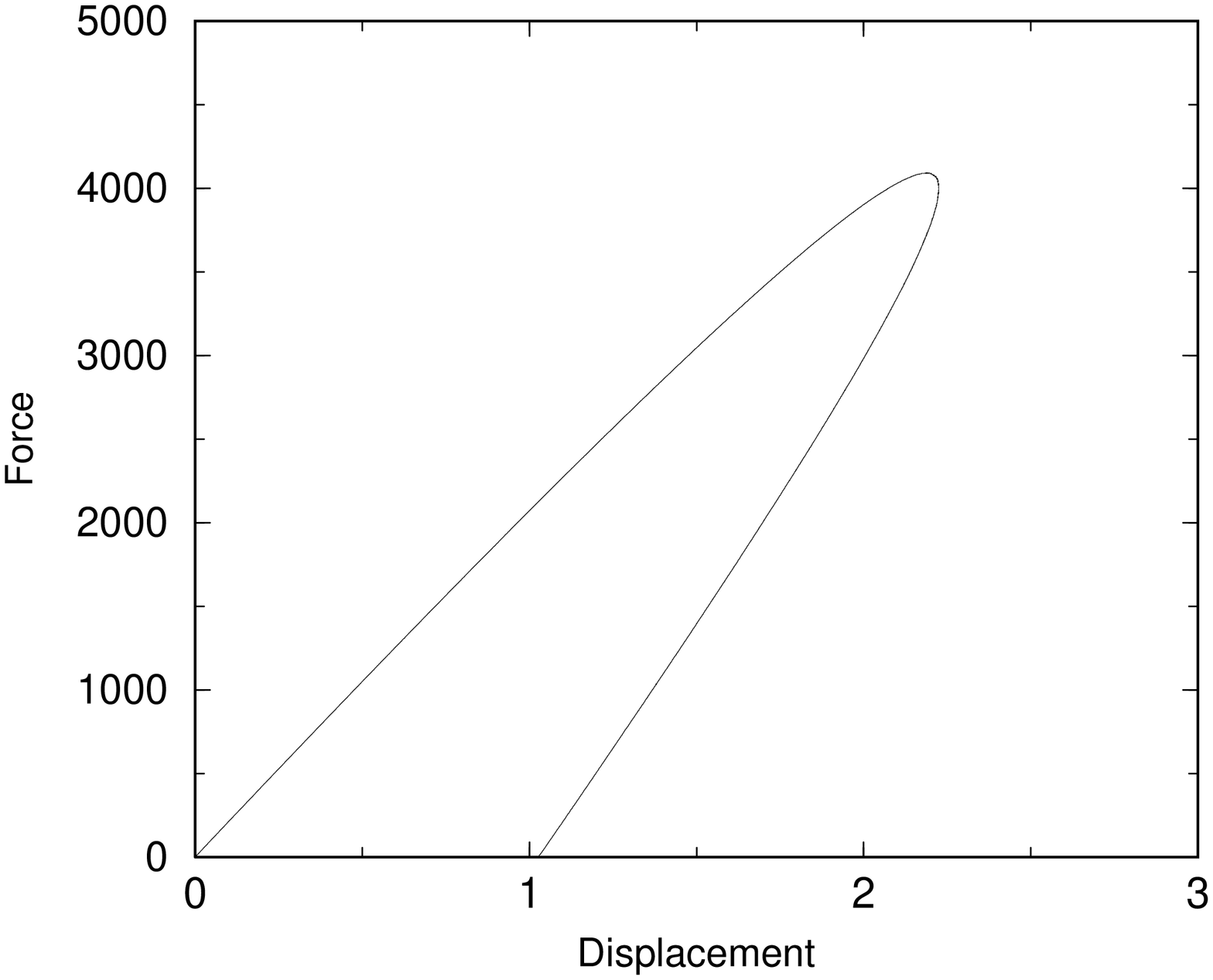}\hskip.1in\includegraphics[%
  width=2.5in,
  height=2.2in,
  angle=-90]{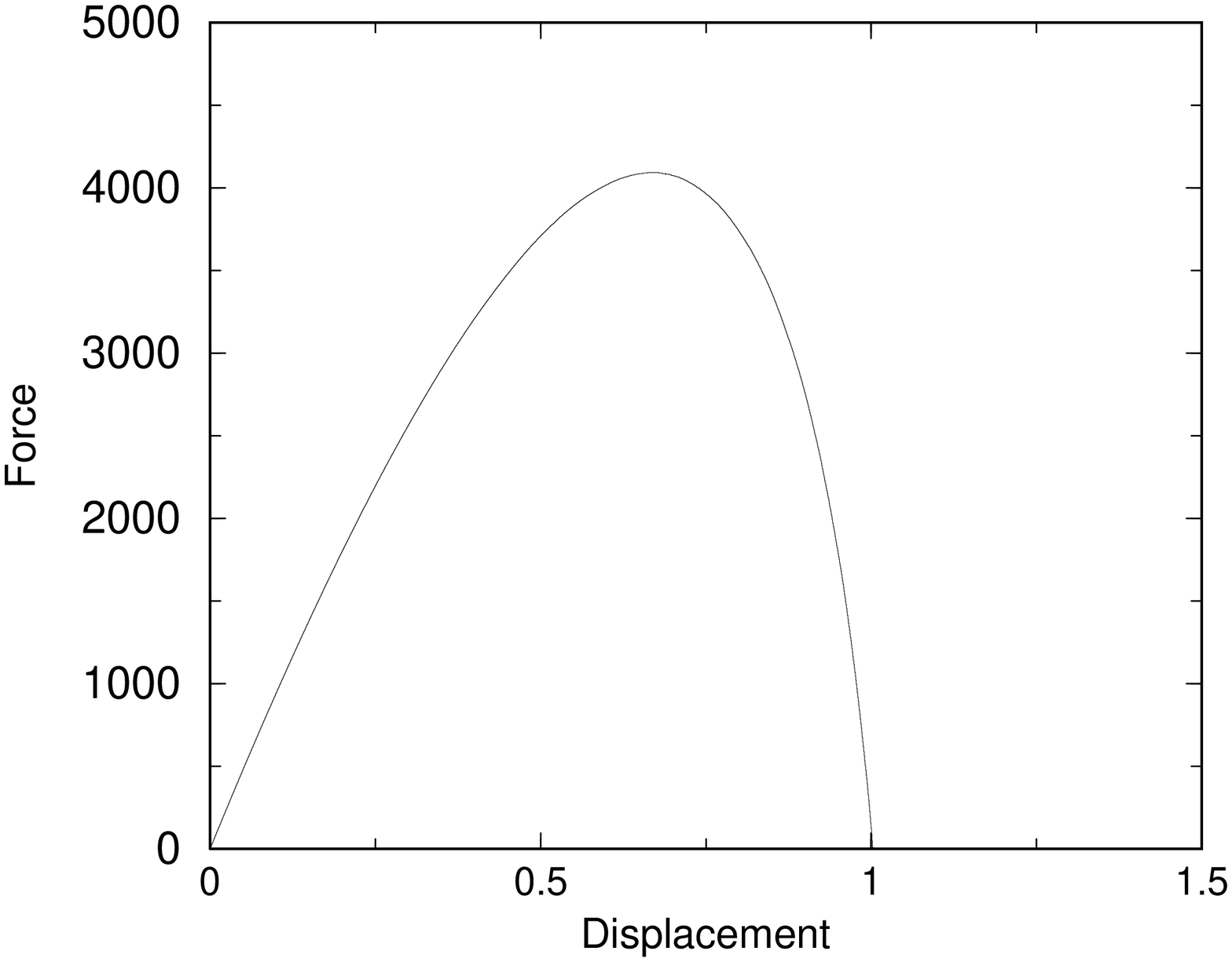}

\vskip.2in

\noindent {\small Fig.\ 13. Force-displacement curve, $128\times 128$ systems with  $e=10$ (left) and  $e=100$ (right).}\\

We now turn to the study of the burst distribution.
We show in Fig. 14 the burst distributions for
$e=10$ and $e=100$. In both cases we find that the burst distribution\index{burst distribution}
follows a power law\index{power-law} with an exponent $\xi=2.6\pm0.1$.
It was argued in Ref.\ \cite{HHP-bhs02} that the value of $\xi$ in this case
is indeed 5/2 as in the global-load-sharing model.

\includegraphics[%
  width=2.5in,
  height=2.2in,
  angle=-90]{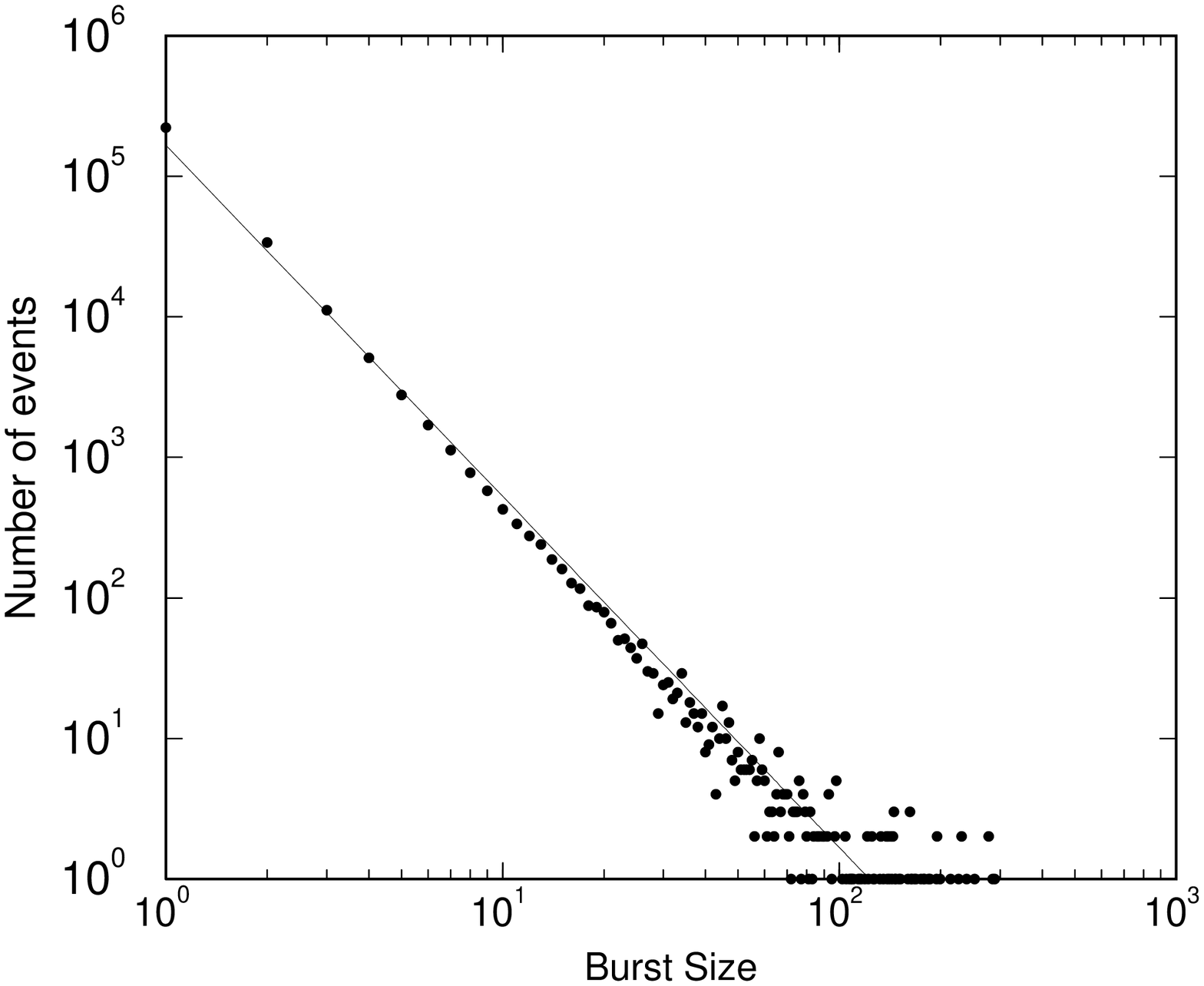}\hskip.1in\includegraphics[%
  width=2.5in,
  height=2.2in,
  angle=-90]{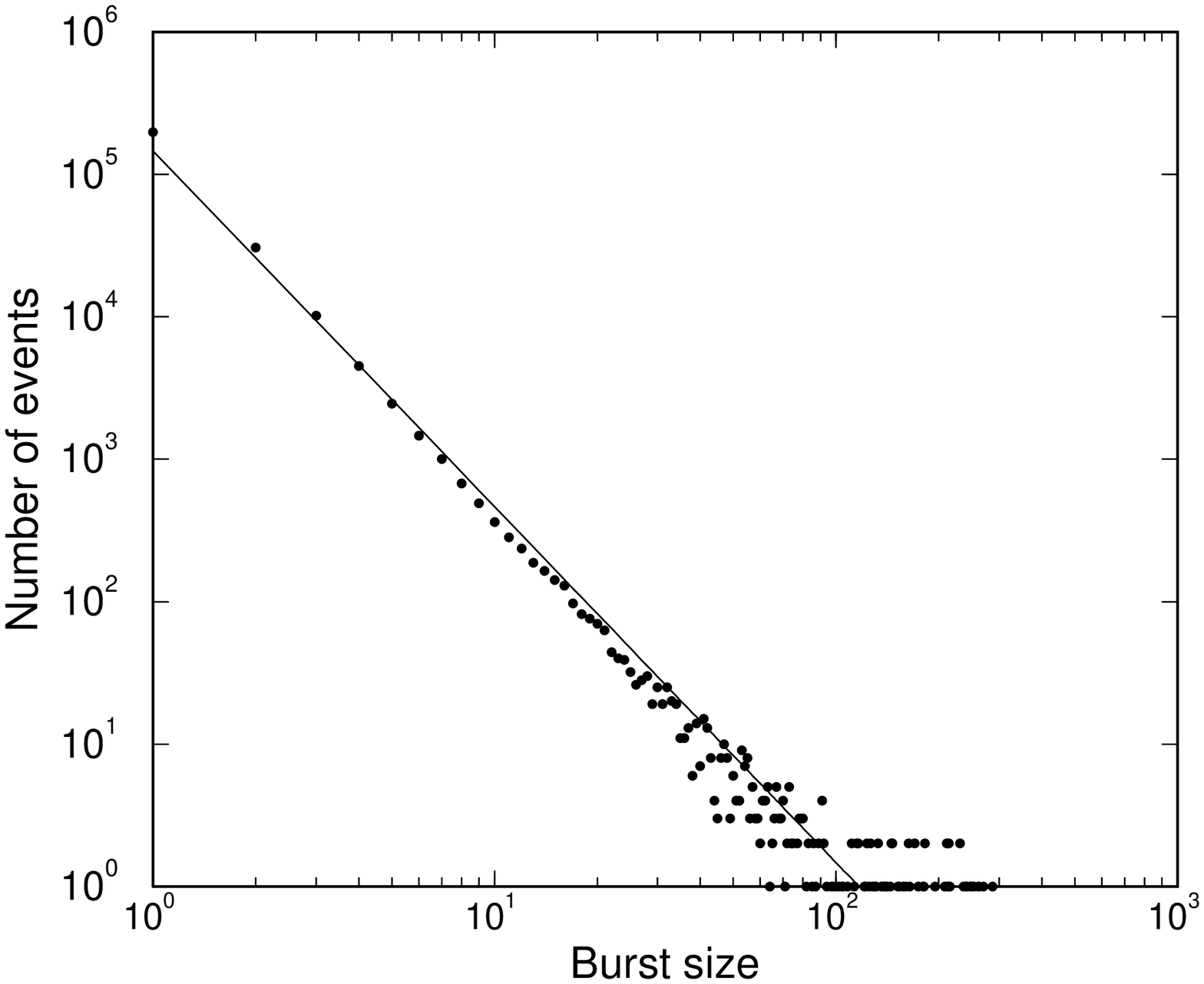}

\vskip.2in

\noindent {\small Fig.\ 14. Burst distribution for $128\times 128$, for $e=10$ (left) and  $e=100$ (right).  The slope of the straight lines is  $-2.5$. }\\

As the failure process proceeds, there is an increasing competition
between global failure due to stress enhancement and local failure due
to local weakness of material.  When the displacement, $D$, is the
control parameter and $e$ is sufficiently small (for example $e=10$),
catastrophic failure eventually occurs due to localization\index{localization}. The onset
of this localization, {\it i.e.\/} the catastrophic regime, occurs when
the two mechanisms are equally important. This may be due to 
self organization \cite{HHP-btw87} occuring at this
point.  In order to test whether this is the case, Batrouni et al.\
\cite{HHP-bhs02} measured
the size distribution of broken bond clusters at the point when $D$
reaches its maximum point on the $F-D$ characteristics, {\it i.e.} the
onset of localization and catastrophic failure.  The analysis was
performed using a Hoshen-Kopelman algorithm \cite{HHP-sa94}.  The result is
shown 
in Fig. 15, for 56 disorder realizations, $L=128$ and
$e=10$. The result is consistent with a power law distribution with
exponent $-1.6$, and consequently with self organization.  If this process
were in the universality class of percolation\index{percolation}, the exponent 
would have been
$-2.05$.  Hence, we are dealing with a new universality class\index{universality class} in this system.

\begin{center}\includegraphics[%
  width=3.0in,
  height=3.0in,
  angle=-90]{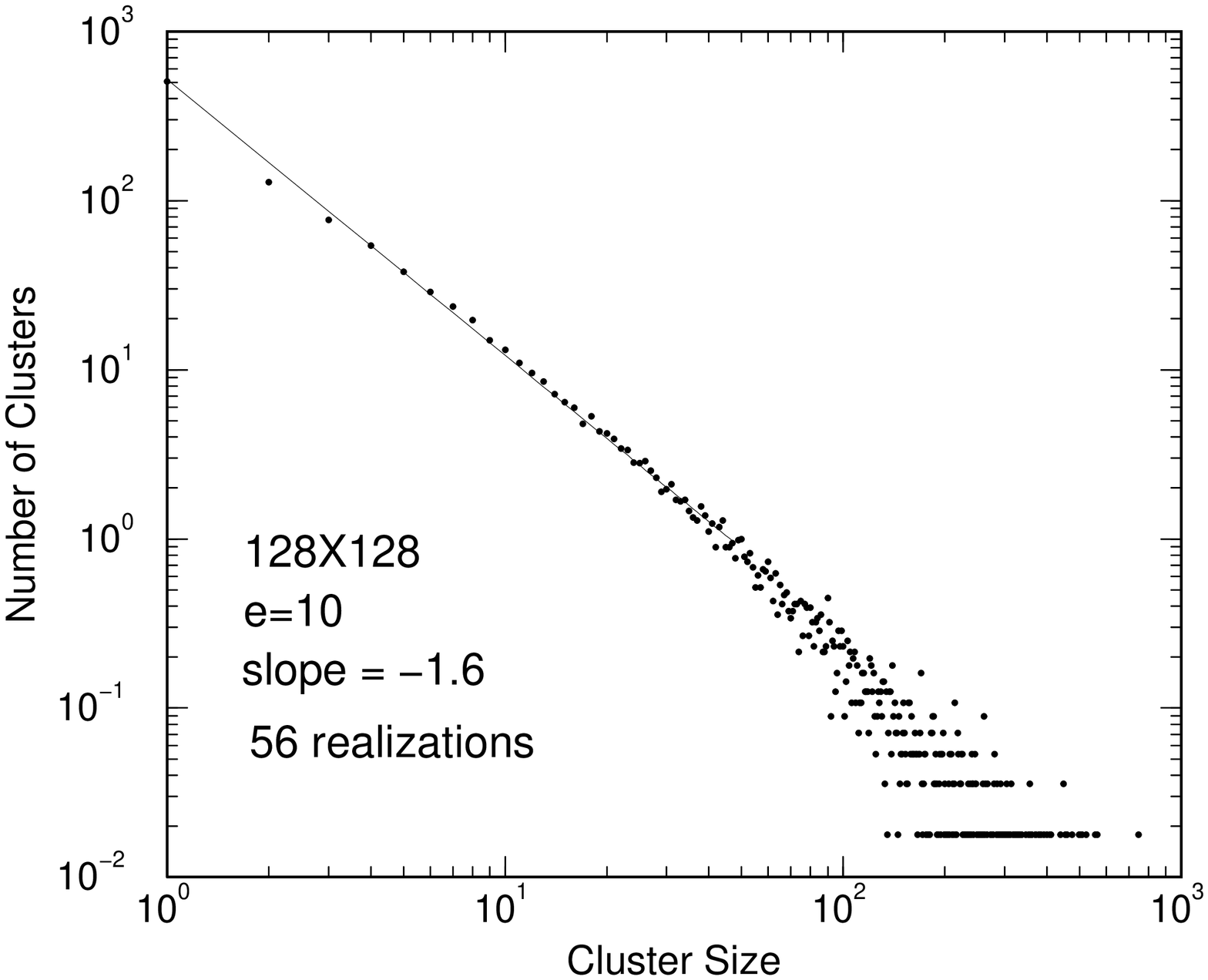}\end{center}

{\small Fig.\ 15. Area distribution of zones where glue has failed for systems 
of size $128\times 128$ and elastic constant $e=10$.  The straight line
is a least square fit and indicates a power law with exponent $-1.6$.}\\
\vskip.3in
\textbf{Acknowledgment:} S. P. thanks NFR (Research Council of Norway) for financial support through Grant No. 166720/V30.

\printindex

\begin{thebibliography}{10}
\bibitem{HHP-Herrmann}H.\ J.\ Herrmann and S.\ Roux, eds.\ {}\emph{Statistical Models
for the Fracture of Disordered Media\/{}} (North-Holland, Amsterdam,
1990). 
\bibitem{HHP-Chakrabarti}B.\ K. Chakrabarti and L.\ G.\ Benguigui \emph{Statistical Physics and Breakdown in Disordered Systems\/{}} (Oxford University Press, Oxford, 1997). 
\bibitem{HHP-Sornette} D.\ Sornette \emph{Critical Phenomena in Natural Sciences\/{}} (Springer Verlag, Berlin, 2000). 
\bibitem{HHP-Daniels}H.\ E.\ Daniels, Proc.\ Roy.\ Soc.\ London \textbf{A183}, 405
(1945). 
\bibitem{HHP-Hemmer}P.\ C.\ Hemmer and A.\ Hansen, ASME J.\ Appl.\ Mech.\ {}\textbf{59}, 909 (1992). 
\bibitem{HHP-Peirce}F.\ T.\ Peirce, J.\ Text.\ Ind.\ {}\textbf{17}, 355 (1926). 
\bibitem{HHP-Kloster}M.\ Kloster, A.\ Hansen, and P.\ C.\ Hemmer, Phys.\ Rev.\ E,
\textbf{56}, 2615 (1997). 
\bibitem{HHP-Hansen2}A.\ Hansen and P.\ C.\ Hemmer, Trends in Statistical Physics \textbf{1}, 213 (1994). 
\bibitem{HHP-Lee}W.\ Lee, Phys.\ Rev.\ B \textbf{50}, 3797 (1994). 
\bibitem{HHP-Pradhan1}S.\ Pradhan, A.\ Hansen, and P.\ C.\ Hemmer, Phys.\ Rev.\ Lett. \textbf{95}, 125501 (2005). 
\bibitem{HHP-Pradhan2}S.\ Pradhan, A.\ Hansen, and P.\ C.\ Hemmer, submitted
to Phys.\ Rev.\ E, cond-mat/0512015 (2005). 
\bibitem{HHP-sornette-sweep}D.\ Sornette, J.\ Phys.\ I France \textbf{4}, 209 (1994). 
\bibitem{HHP-Zapperi-crossover}S.\ Zapperi, P.\ K.\ V.\ V.\ Nukula, and S.\ Simunovic, Phys.\ Rev.\ E.\ {}\textbf{71}, 026106 (2005). 
\bibitem{HHP-Kawamura}H.\ Kawamura, arXiv:cond-mat/0603335, (2006).
\bibitem{HHP-Silveira}R.\ de Silveira, Am.\ J.\ Phys. \textbf{67}, 1177 (1999).
\bibitem{HHP-Pradhan4}S.\ Pradhan and B.\ K.\ Chakrabarti, Phys.\ Rev.\ E \textbf{65}, 016113 (2001). 
\bibitem{HHP-Pradhan5}S.\ Pradhan, P.\ Bhattacharyya, and B.\ K.\ Chakrabarti, Phys.\ Rev.\ E \textbf{66} 016116 (2002). 
\bibitem{HHP-Pradhan6}P.\ Bhattacharyya, S.\ Pradhan, and B.\ K.\ Chakrabarti, Phys.\ Rev.\ E \textbf{67}, 046122 (2003). 
\bibitem{HHP-Zapperi}S.\ Zapperi, P.\ Ray, H.\ E.\ Stanley, and A.\ Vespignani, Phys.\ Rev.\ Lett.\ {}\textbf{85}, 2865 (2000). 
\bibitem{HHP-Silveira}R.\ de Silveira, Am.\ J.\ Phys. \textbf{67}, 1177 (1999).
\bibitem{HHP-Harlow}D.\ G.\ Harlow, Proc.\ Roy.\ Soc.\ Lond.\ Ser.\ A \textbf{397},
211 (1985). 
\bibitem{HHP-Harlow2}D.\ G.\ Harlow and S.\ L.\ Phoenix, J.\ Mech.\ Phys.\ Solids
\textbf{39}, 173 (1991). 
\bibitem{HHP-Duxbury}P.\ M.\ Duxbury and P.\ M.\ Leath, Phys.\ Rev.\ B \textbf{49},
12676 (1994). 
\bibitem{HHP-Harlow3}D.\ G.\ Harlow and S.\ L.\ Phoenix, Int.\ J.\ Fracture \textbf{17},
601 (1981). 
\bibitem{HHP-Phoenix}S.\ L.\ Phoenix and R.\ L.\ Smith, Int.\ J.\ Sol.\  Struct.\ {}\textbf{19},
479 (1983). 
\bibitem{HHP-Kuo}C.\ C.\ Kuo and S.\ L.\ Phoenix, J.\ Appl.\ Prob.\ {} \textbf{24},
137 (1987). 
\bibitem{HHP-Hansen}A.\ Hansen and P.\ C.\ Hemmer, Phys.\ Lett.\ A \textbf{184},
394 (1994). 
\bibitem{HHP-Ding}S.\ D.\ Zhang and E.\ J.\ Ding, Phys.\ Lett.\ A \textbf{193}
425 (1994). 
\bibitem{HHP-Smith}R.\ L.\ Smith, Ann.\ Prob.\ {}\textbf{10}, 137 (1982). 
\bibitem{HHP-Zhang}S.\ D.\ Zhang and E.\ J.\ Ding, Phys.\ Rev.\ B \textbf{53},
646 (1996). 
\bibitem{HHP-Zhang2}S.\ D.\ Zhang and E.\ J.\ Ding, J.\ Phys.\ A \textbf{28}, 4323
(1995). 
\bibitem{HHP-Pradhan3}S.\ Pradhan, B.\ K.\ Chakrabarti, and A.\ Hansen, Phys.\ Rev.\
E \textbf{71}, 036149 (2005). 
\bibitem{HHP-Hidalgo}R.\ C.\ Hidalgo, Y.\ Moreno, F.\ Kun, and H.\ J.\ Herrmann,
Phys.\ Rev.\ E \textbf{65}, 046148 (2002). 
\bibitem{HHP-bhs02}  G.\ G.\ Batrouni, A.\ Hansen, and J.\ Schmittbuhl,
Phys. Rev. E {\bf 65}, 036126 (2002).

\bibitem{HHP-ll58} L.\ Landau and E.\ M.\ Lifshitz, {\sl Theory of Elasticity\/}
(Clarendon Press, Oxford, 1958).

\bibitem{HHP-j85} K.\ L.\ Johnson, {\sl Contact Mechanics\/} (Cambridge University
Press, Cambridge, 1985).

\bibitem{HHP-bh88} G.\ G.\ Batrouni and A.\ Hansen, J.\ Stat.\ Phys.\ {\bf 52},
747 (1988).

\bibitem{HHP-ptvf92} W.\ H.\ Press, S.\ A.\ Teukolsky, W.\ T.\ Vetterling, and
B.\ P.\ Flannery, {\sl Numerical Recipes in Fortran 77: The Art of
Scientific Computing\/} (Cambridge University Press, Cambridge, 1992).

\bibitem{HHP-bhn86} G.\ G.\ Batrouni, A.\ Hansen, and M.\ Nelkin, Phys.\
Rev.\ Lett.\ {\bf 57}, 1336 (1986).

\bibitem{HHP-btw87} P.\ Bak, C.\ Tang and K.\ Wiesenfeld, Phys.\ Rev.\ Lett.\
{\bf 59}, 381 (1987).

\bibitem{HHP-sa94} D.\ Stauffer and A.\ Aharony, {\it Introduction to Percolation
Theory\/} (Taylor and Francis, London, 1992).

 
\end{thebibliography}
\end{document}